\documentclass[10pt,journal,comsoc]{IEEEtran}
\usepackage{url}
\usepackage{caption}
\usepackage[noadjust]{cite}
\usepackage{graphicx}
\usepackage{multirow}
\usepackage{comment}
\usepackage{makecell}
\usepackage{ragged2e}
\usepackage{rotating}
\usepackage{lscape}
\usepackage{nomencl}
\usepackage{dblfloatfix}
\makenomenclature
\usepackage{wrapfig}
\usepackage{makecell}
\usepackage{balance}
\usepackage[utf8]{inputenc}
\usepackage{ragged2e}
\usepackage{booktabs}

\usepackage{pifont}
\usepackage{tabulary}
\usepackage[table, svgnames]{xcolor}
\usepackage{lipsum}

\usepackage{array, tabularx, multirow, booktabs, diagbox}
\usepackage{adjustbox}

\newcommand{\cmark}{\ding{51}}%
\newcommand{\xmark}{\ding{55}}%
\ifCLASSOPTIONcompsoc
\else
\fi
\ifCLASSINFOpdf
\else
\fi

\begin{document}
\title{Digital Twin of Wireless Systems: Overview, Taxonomy, Challenges, and Opportunities}


\author{Latif~U.~Khan,~Zhu~Han,~\IEEEmembership{Fellow,~IEEE},~Walid~Saad,~\IEEEmembership{Fellow,~IEEE},~Ekram~Hossain,~\IEEEmembership{Fellow,~IEEE},~Mohsen~Guizani,~\IEEEmembership{Fellow,~IEEE},~Choong~Seon~Hong,~\IEEEmembership{Senior~Member,~IEEE}

\IEEEcompsocitemizethanks{
\IEEEcompsocthanksitem L.~U.~Khan~and~C.~S.~Hong are with the Department of Computer Science \& Engineering, Kyung Hee University, Yongin-si 17104, South Korea.

\IEEEcompsocthanksitem Zhu Han is with the Electrical and Computer Engineering Department, University of Houston, Houston, TX 77004 USA, and also with the Computer Science Department, University of Houston, Houston, TX 77004 USA, and the Department of Computer Science and Engineering, Kyung Hee University, South Korea.

\IEEEcompsocthanksitem Walid Saad is with the  Wireless@VT, Bradley Department of Electrical and Computer Engineering, Virginia Tech, Blacksburg, VA 24061 USA.

\IEEEcompsocthanksitem M. Guizani is with the Department of Computer Science and Engineering at Qatar University, Doha, Qatar.

\IEEEcompsocthanksitem E. Hossain is with Department of Electrical and Computer Engineering at University of Manitoba, Winnipeg, Canada.

}}

\markboth{}{}%

\maketitle






\begin{abstract} 
Future wireless services must be focused on improving the quality of life by enabling various applications, such as extended reality, brain-computer interaction, and healthcare. These applications have diverse performance requirements (e.g., user-defined quality of experience metrics, latency, and reliability) that are challenging to be fulfilled by existing wireless systems. To meet the diverse requirements of the emerging applications, the concept of a digital twin has been recently proposed. A digital twin uses a virtual representation along with security-related technologies (e.g., blockchain), communication technologies (e.g., $6$G), computing technologies (e.g., edge computing), and machine learning, so as to enable the smart applications. In this tutorial, we present a comprehensive overview on digital twins for wireless systems. First, we present an overview of fundamental concepts (i.e., design aspects, high-level architecture, and frameworks) of digital twin of wireless systems. Second, a comprehensive taxonomy is devised for both different aspects. These aspects are \textit{twins for wireless} and \textit{wireless for twins}. For the \textit{twins for wireless} aspect, we consider parameters, such as twin objects design, prototyping, deployment trends, physical devices design, interface design, incentive mechanism, twins isolation, and decoupling. On the other hand, for \textit{wireless for twins}, parameters such as, twin objects access aspects, security and privacy, and air interface design are considered. Finally, open research challenges and opportunities are presented along with causes and possible solutions.      
\end{abstract}



\section{Introduction}
\setlength{\parindent}{0.7cm}Emerging Internet of Everythings (IoE) applications, such as haptics, brain-computer interaction, flying vehicles, and extended reality (XR), among others, will enable a merger of digital and physical worlds \cite{al2015internet,atzori2010internet,li20185g,ge2018big}. These IoE applications have widely diverse requirements (e.g., user experience, reliability, and latency). To meet these diverse requirements, there is a need to assist the wireless systems by novel technologies. These new technologies will enable the wireless systems to meet the diverse requirements via enabling the two main trends: \textit{self-sustaining wireless systems} and \textit{proactive-online-learning-enabled systems} \cite{khan2021digital}. Generally, wireless systems rely on intelligent, seamless, and ubiquitous connectivity for meeting the diverse requirements of end-users. To enable wireless systems with these features, there is a need for self-sustaining wireless systems. Self-sustaining wireless systems will offer efficient management of wireless systems with minimum possible intervention from the network operators/end-users. Such self-sustaining wireless systems can use optimization theory, game theory, and graph theory, among others to enable IoE services. On the other hand, IoE services have highly dynamic requirements in terms of user-defined metrics, latency, data rate, and reliability, among others. To meet these highly dynamic and extreme requirements, there is a need to efficiently enable interaction between various players of wireless systems. These players are security-related technologies (e.g., blockchain), computing technologies (e.g., edge computing), and wireless channel resources (e.g., terahertz band, millimeter wave). Therefore, upon request from the end-users, there is a need to provide them instant services especially for strict latency applications (e.g., extended reality). To enable such kinds of instant services to end-users, we must perform intelligent analytics (i.e., proactive learning) prior to service request for efficient resource management. Therefore, it is necessary to propose proactive-online learning-based wireless systems.        
  \par
  
\begin{figure*}[!t]
	\centering
	\captionsetup{justification=centering}
	\includegraphics[width=18cm, height=15.5cm]{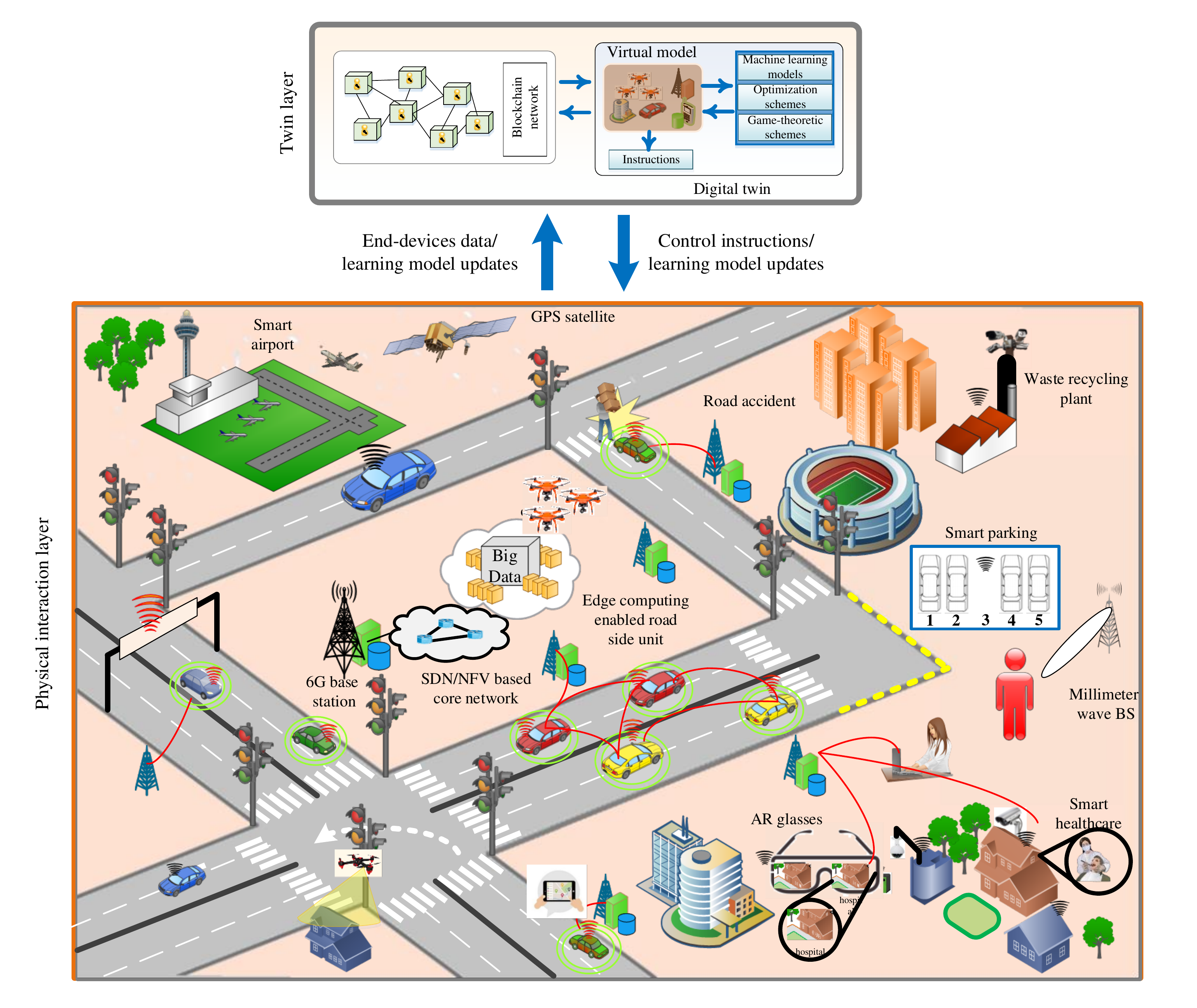}
	\caption{Conceptual overview of a digital twin for wireless systems.}
	\label{fig:twinoverview}
\end{figure*}
To design a wireless system by following the aforementioned trends of self-sustaining wireless systems and proactive-online-learning-enabled systems, we can create a digital twin to represent the wireless system \cite{khan2021digital}. A digital twin will use a virtual representation of the physical system to enable IoE applications \cite{khan2021digital,tao2018digital,haag2018digital}. In detail, a digital twin-based wireless system will use optimization theory, game theory, and machine learning in addition to the virtual representation of a wireless system. Additionally, to enable transparent and immutable handling of data, a digital twin-based system will use blockchain. An overview of digital twin-based wireless systems is given in Fig.~\ref{fig:twinoverview}. We can divide the twin-based wireless systems architecture into physical interaction layer and twins layer. The physical interaction layer covers all the physical objects necessary for a given wireless system application, such as end-devices, edge/cloud servers, base stations, and core network elements, among others. On the other hand, the twin layer is a logical layer that contains logical twin objects. More detailed discussion about the creation of twin objects will be provided in Sections~\ref{architecture} and \ref{trends}). Next, we discuss research trends and statistics for digital twins. \par

\subsection{Market Statistics and Research Trends}
The IoT market will grow at a Compound Annual Growth Rate (CAGR) of $26.9\%$ during the period of $2017-2022$ \cite{IoT_tastiscts_1}. The market share will grow from $170.6$ Billion USD in $2017$ to $561.0$ Billion USD in $2022$. The main causes of an increase in the IoT market are smart buildings, smart grids, smart industries, intelligent transportation. The key players of IoT markets are General Electric (US), Bosch Software Innovation GmbH (Stuttgart, Germany), Amazon Web Services Inc. (US), Hewlett-Packard Enterprise (US), Google Inc. (US), PTC Inc. (US),  International Business Machine (IBM) Corporation (US),  Oracle Corporation (US),  Microsoft Corporation (US),  Cisco Systems, Inc. (US),  SAP SE (Walldorf, Germany), and Intel Corporation (US). Among different regions, such as Latin America, MEA, APAC, Europe, and North America, it is expected that China will lead the IoT market in APAC region \cite{IoT_tastiscts_2}. Globally, the US will have the highest share in IoT markets by $2029$ \cite{IoT_tastiscts_3} and will be followed by China, Japan, and Germany. \par        	
According to statistics, the market of digital twins will grow at a CAGR rate of $58\%$ during the period of $2020-2026$ \cite{Twin_tastiscts_1}. The market value of digital twin in $2020$ was $3.1$ Billion USD and it will reach $48.2$ Billion USD by $2026$. The key factors in the rise of the digital twin market include the rise in the demand for manufacturing monitoring assets, intelligent analytics-based healthcare systems, smart warehouse, and intelligent transportation, among others \cite{Twin_tastiscts_1, Twin_tastiscts_2}. Due to the increasing importance of digital twin in the development of smart applications, various countries, such as Brazil, Norway, Mexico, China, Japan, South Korea, and Singapore, are trying to implement twin based systems. The key market players of digital twin market are SWIM.AI (USA), Robert Bosch (Germany), ANSYS (USA), Siemens AG (USA), Oracle (USA), SAP (Germany), Microsoft Corporation (USA), PTC (USA), IBM (USA), and General Electric (USA). Additionally, system digital twin among all kinds, such as product digital twin, process digital twin, and system digital twin will have the highest market share by $2026$. From the aforementioned discussion, it is clear that both digital twin and IoT will serve as one of the promising areas for future research. \par

\begin{table*}[]
\rowcolors{2}{gray!25}{white}
\caption {Summary of existing surveys and tutorials with their primary focus.} \label{tab:surveyssummaries} 
\begin{center}
\begin{tabular}{p{2cm}p{2cm}p{2cm}p{2cm}p{6cm}}
\toprule 
    \textbf{Reference}   & \textbf{Wireless for twins}& \textbf{Twins for wireless}  & \textbf{Taxonomy}&\textbf{Remark} \\ \midrule
    Minerva \textit{et al.}~\cite{twin_pro_ieee} &\xmark &\xmark & \xmark & This survey comprehensively presents technical features, scenarios and architectural models for digital twin in the context of IoT. \\ \midrule
    Wu \textit{et al.}~\cite{wu2021digital} & \cmark & \xmark& \xmark& This survey presents the twin fundamentals, as well as key enabling technologies, and open challenges.\\ \midrule
    Barricelli \textit{et al.}~\cite{barricelli2019survey} & \xmark & \xmark & \xmark & This paper presents the fundamentals, implementation details, and applications of digital twin.\\ \midrule
    Yaqoob \textit{et al.}~\cite{yaqoob2020blockchain} & \xmark & \xmark& \xmark & The authors discussed the role of blockchain towards enabling digital twin. Additionally, taxonomy was also devised.\\ \midrule 
     Suhail \textit{et al.}~\cite{suhail2021blockchain} & \xmark & \xmark & \xmark & Blockchain-based twins are discussed. Furthermore, research trends and future directions were also presented.\\ \midrule
    Khan \textit{et al.}~\cite{khan2021digital} & \xmark & \cmark &\xmark & This tutorial presents the key design requirements, architectural trends, and future directions for digital-twin-enabled $6$G.\\ \midrule
    Our Tutorial & \cmark  & \cmark   &\cmark   & N.A\\ 
\bottomrule 
\end{tabular}
\end{center}
\end{table*}

\subsection{Existing Surveys and Tutorials}
Few surveys and tutorials have reviewed digital twins \cite{twin_pro_ieee, wu2021digital, barricelli2019survey, yaqoob2020blockchain, suhail2021blockchain,khan2021digital}. The authors in \cite{twin_pro_ieee} focused on digital twin in the context of IoT. They discussed the digital twin concept with architectural elements as well as key enablers. Another work \cite{wu2021digital} surveyed the key technologies along with the use case of digital twins towards enabling IoT applications. Barricelli \textit{et al.} in \cite{barricelli2019survey} presented the key concepts, applications, and design implications. Furthermore, they presented few open research challenges. On the other hand, the works in \cite{yaqoob2020blockchain} and \cite{suhail2021blockchain} focused mainly on the role of blockchain in enabling digital twins. The authors in \cite{yaqoob2020blockchain} presented the key benefits of using blockchain for digital twins. Additionally, they devised taxonomy and presented a few open challenges. Suhail \textit{et al.} in \cite{suhail2021blockchain} systematically reviewed the role of blockchain in enabling digital twins. The work in \cite{khan2021digital} presented the role of digital twin towards enabling of $6$G wireless system. Additionally, the authors provided architectural trends for twin-based wireless systems. Different from the existing surveys and tutorials on digital twins \cite{twin_pro_ieee, wu2021digital, barricelli2019survey, yaqoob2020blockchain, suhail2021blockchain,khan2021digital}, the goal of our survey is to comprehensively discuss the fellowship of digital twin and wireless systems, as given in Table~\ref{tab:surveyssummaries}. We present the fundamentals of digital twins and derive a general definition in the context of wireless systems. A general architecture along with design aspects is presented. We also derive a comprehensive taxonomy of digital twins-based wireless systems and presented open research challenges.  \par     

\begin{figure}[!t]
	\centering
	\captionsetup{justification=centering}
	\includegraphics[width=8cm, height=15.5cm]{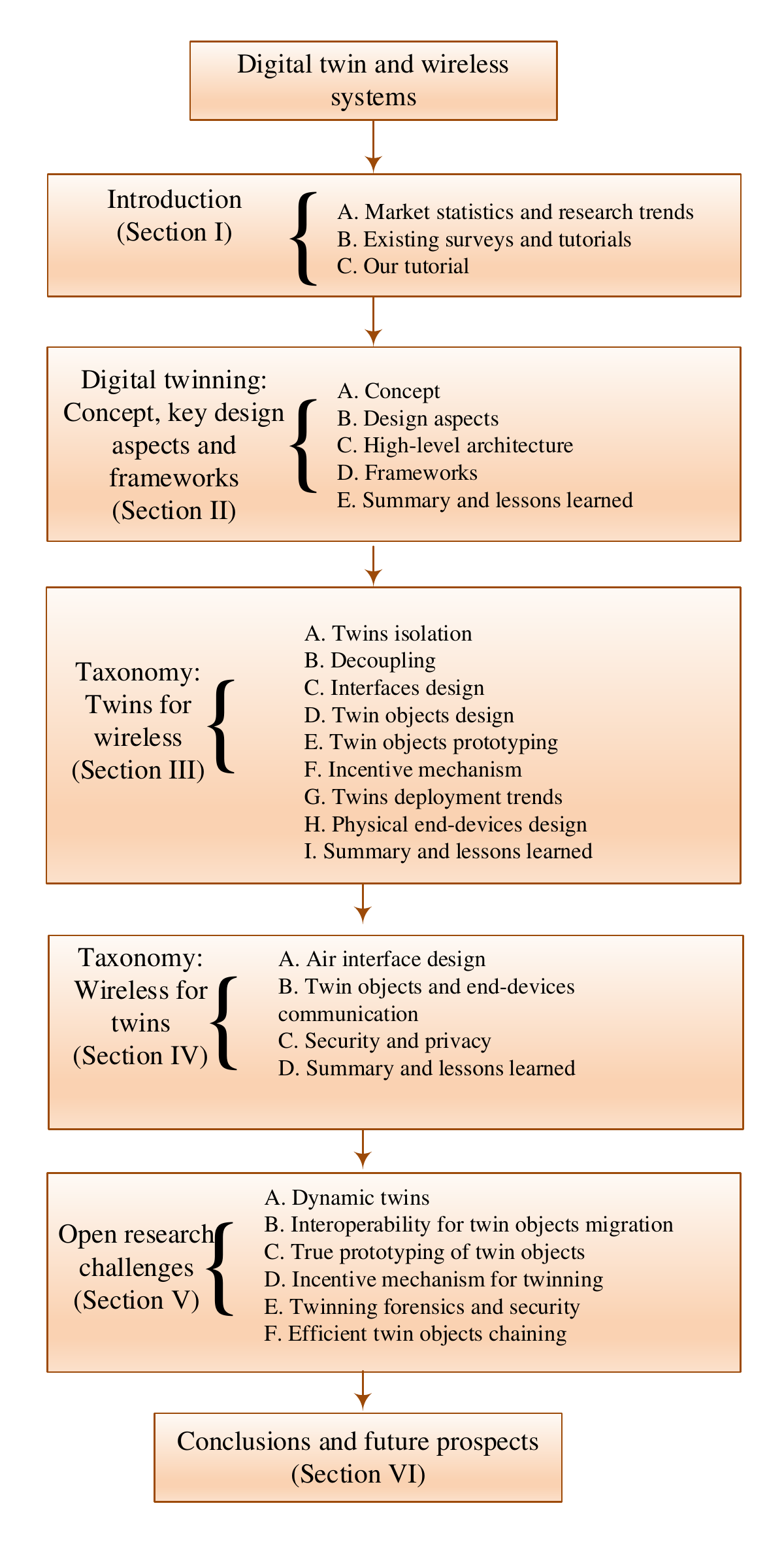}
	\caption{Organization of the tutorial.}
	\label{roadmap}
\end{figure}

\subsection{Our Tutorial}
Our tutorial (see organization in Fig.~\ref{roadmap}) aims to answer the following questions:\par
\begin{itemize}
    \item What is a digital twin in the context of wireless systems? What are the main design aspects of digital twin of wireless systems?
    \item How does one use digital twin for enabling wireless systems? What are the challenges for digital twin signaling over the wireless channel?  
    \item How can we classify the research areas for digital twin of wireless system? 
    \item What are the open research challenges and their preferable solutions in enabling the digital twins of wireless systems?
\end{itemize}
The main focus of our tutorial is to consider two main aspects: digital twins for wireless and wireless for twins. Twin for wireless deals with the role of the digital twin in enabling wireless systems. Wireless for twins deals with the efficient utilization of wireless resources for enabling effective twin signalings over a wireless link. Different from the existing works \cite{twin_pro_ieee, wu2021digital, barricelli2019survey, yaqoob2020blockchain, suhail2021blockchain,khan2021digital}, we present a detailed concept, key design aspects, and high-level architecture of digital twins for wireless systems. We also present a comprehensive taxonomy that covers both twins for wireless and wireless for twins aspects. Furthermore, novel open research challenges with their causes and solutions are provided. Our contributions are summarized as follows.\par
\begin{itemize}
    \item We present a concept, key design aspects, and high-level architecture for digital twin of wireless systems.
    \item A comprehensive taxonomy considering both twins for wireless and wireless for twins, is provided. We consider twin objects design, twin objects prototyping, twin objects deployment trends, interface design, incentive mechanism, twin objects access aspects, twins isolation, decoupling, security and privacy, and AI-enabled air interface design, as parameters. 
    \item Several open challenges are presented. Moreover, promising solutions are also provided.
\end{itemize}




\section{Digital Twins: Concept, Key Design Aspects, and Frameworks}
\subsection{Concept}
A digital twin is a virtual representation of the physical system serving as a digital counterpart \cite{Twin_Def_1, Twin_Def_2}. The main purpose of a digital twin is to jointly optimize the cost and performance of the overall process using various emerging technologies (e.g., virtual modeling, simulation technologies, blockchain, edge computing, cloud computing) and optimization tools (e.g., machine learning, game theory, graph theory). Digital twin provides proactive analysis of the physical process using various simulation tools (e.g., AMEsim, SimScale, Simulink \cite{Twin_soft_1, Twin_soft_2}), artificial intelligence, mathematical optimization, game theory, and graph theory, among others \cite{khan2021digital}. Such analyses enable us to optimize the overall design. Digital twin technology has gained significant interest since $2002$ when the firm, namely, \textit{Challenge Advisory} hosted a presentation at University of Michigan \cite{Twin_history_1}. They discussed the fundamental elements of a digital twin, such as virtual space, real space, and the information flow between them. Prior to the event at the University of Michigan, the US space agency NASA proposed the use of a digital twin around $1960$'s for analyzing the space systems at the ground. \par
\begin{figure}[!t]
	\centering
	\captionsetup{justification=centering}
	\includegraphics[width=8cm, height=7.5cm]{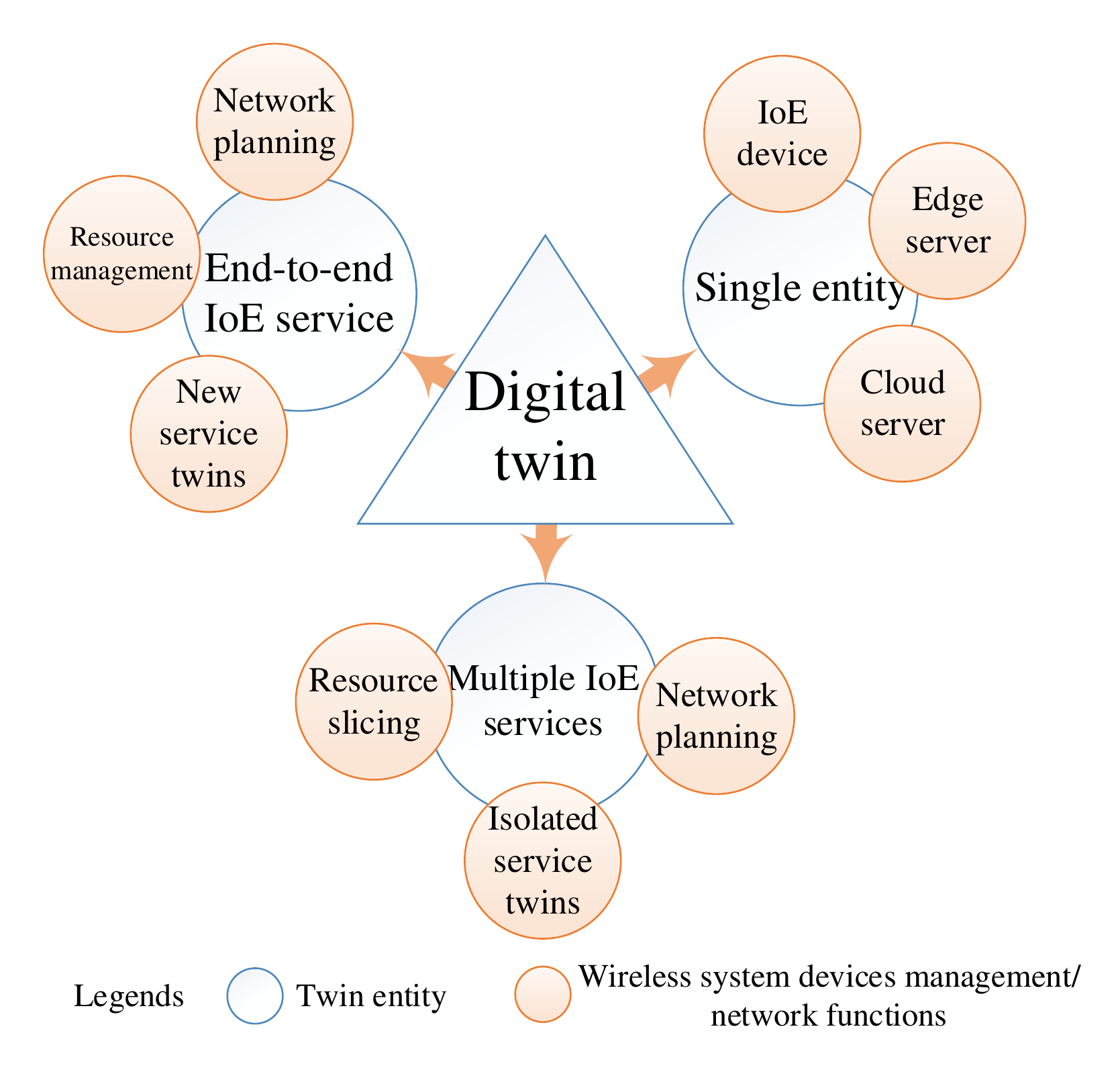}
	\caption{Conceptual overview of a digital twin for wireless systems.}
	\label{fig:twindefinition}
\end{figure}

Digital twins are categorized by various sources in many ways, as given in Table~\ref{tab:twin categories}. According to Siemens, a digital twin can be divided into product digital twins, production digital twins, and performance digital twins \cite{Twin_type_1}. Furthermore, the twins can be categorized into status twin, simulation twin, and operational twin \cite{Twin_type_4}. The classification of twins by different sources is based on their objective and coverage (i.e, single entity or whole system). From all the types of twins \cite{Twin_type_1, Twin_type_2, Twin_type_3, Twin_type_4}, we generalize the definition and types of a twins for a wireless system, as shown in Fig.~\ref{fig:twindefinition}. Digital twin for $6$G can be intended for a single entity (i.e, edge server management and IoE end-devices management), an end-to-end service (i.e., resource management, new service design, and network planning), and multi-services (i.e., resource slicing for different twins, isolation of different service twins, and network planning). From the aforementioned discussion, one can say that \textit{digital twin for $6$G can enable analysis, design, and real-time monitoring with control of devices to enable IoE services for cost-efficient and resource optimized operation}.

\begin{table*}[!t]
  \centering
  \caption {Digital twin categories: source, type, and explanation.} \label{tab:twin categories} 
  \begin{tabular}{>{\centering\arraybackslash}m{1.5cm}p{4cm}p{6cm}p{4cm}}
    \hline
   
    \multirow{1}{*}{Reference} & \multirow{1}{*}{Category (defined by source)}&  \multirow{1}{*}{Description by the source} &  \multirow{1}{*}{Key objective}  \\ 
    \hline 
    \multirow{3}{*}{Siemens~\cite{Twin_type_1}} 
    & Product digital twins & To use a digital for enabling a new product. This twin will offers us with the ability to make various experiments by varying different system parameters to analyze the cost, efficiency, and performance for a new product before it is actually manufactured. & New product design \\
     \cline{2-4}  
     & Production digital twins & To use a digital twin in process of manufacturing. Digital twins of the various components are tested for performance and cost. This will lead efficient selection/placement of manufacturing process component in real time deployment. & Efficient deployment of system components   \\
     \cline{2-4}
     &Performance digital twins &  To use a digital twin for capturing and acting on operational data. This twin will enable smart plants (e.g., smart grid, smart industry) to operate in an efficient way to optimize the cost and performance by using the generated data. This data can be further used by machine learning models and other techniques to enhance the twin performance. & Data analysis for performance enhancement \\
     \hline
     \multirow{3}{*}{Vercator~\cite{Twin_type_2}} 
     & Standalone digital twins & This twin refers to the virtual models of individual entities (e.g., Heating, ventilation, and air conditioning system in a smart home) of a system. &  Modeling of a single entity. \\
     \cline{2-4}
     & Duplicated digital twins & This twin refers to twin for optimized use of multiple stand-alone twins (e.g., car twins and smart industry twins). &  Performance optimization of a component that is made of many twins in a complex system. \\
     \cline{2-4}
     & Enhanced digital twins & This twin deals with modeling of complex systems that have a broad scale than those of a duplicate twin and will consist of multiple duplicate twins.   & To enhance the performance of a complete system (e.g., 6G-based smart surveillance in a city).    \\
     
     \hline 
          \multirow{3}{*}{Tributech~\cite{Twin_type_3}} 
        & Component twins& This twin represents a single component in an entire system. & Single specific component in a system. \\
        \cline{2-4}
        & Asset twins & It refers to twin that can manage the system that is made up of multiple component twins (e.g., autonomous car engine). & System multiple components. \\
        \cline{2-4}
        & System twins & This twin scales by covering a large system (e.g., entire autonomous car). & Single system   \\
        \cline{2-4}
        & Process twins & This twin models the entire process facility (e.g., autonomous car plant).  & Entire plant\\
        \hline
         \multirow{3}{*}{XMPRO~\cite{Twin_type_4}}
          & Status twin & This twin is used to monitor the system (e.g., dashboards and simple alerting systems). & To monitor the phenomenon. \\
          \cline{2-4}
          & Simulation twin & It enables us analyze the system performance via simulation and provide insights into future states. & To analyze the system performance. \\
          \cline{2-4}
          & Operational twin & It enables us to interact with the system and control the system for efficient and cost performance. & To efficiently control the system.\\
          \hline
                
  \end{tabular}
\end{table*}

\begin{figure}[!t]
	\centering
	\captionsetup{justification=centering}
	\includegraphics[width=8cm, height=5cm]{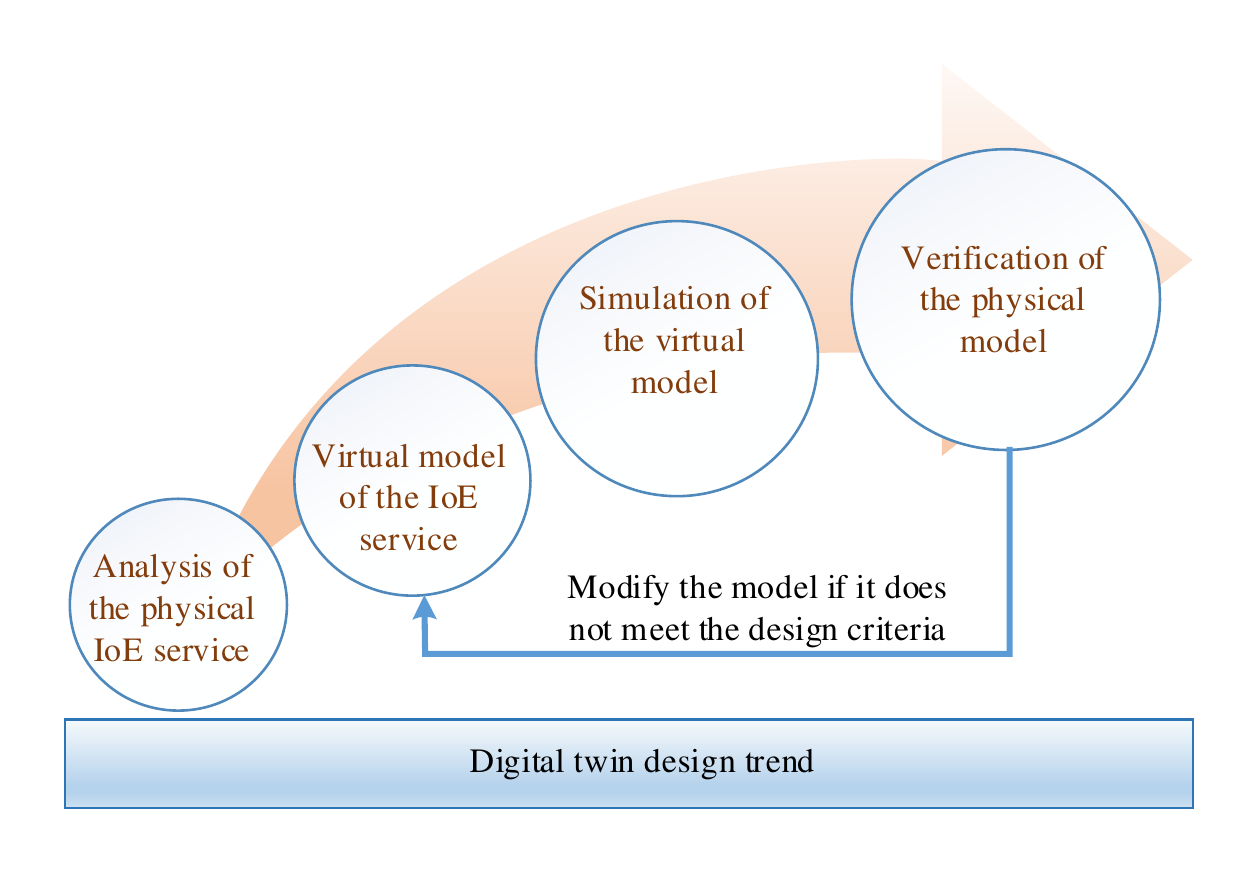}
	\caption{Digital twin design steps.}
	\label{fig:designsteps}
\end{figure}

The implementation process of a digital twin starts with the analysis of a physical system, as shown in Fig.~\ref{fig:designsteps}.  Such analysis gives us insights into the system specifications, inputs, outputs, and environment dynamics. Next to physical system analysis, one can design a virtual system that represents the physical system. Modeling a virtual system requires careful design and requires overcoming many challenges. The challenges are an accurate representation of a system using a mathematical equations or using machine learning models. Additionally, some of the real-time system parameters (e.g., IoE device operating frequency, wireless channel conditions) are very difficult to model accurately due to their uncertain nature with the time. Therefore, one must try to accurately estimate these parameters while performing the simulation of the virtual model. After virtual modeling, there is a need to verify the virtual design by simulating the virtual model and comparing the performance with the physical system. One must modify the virtual model in this phase to make the virtual model closest possible to the physical system. The final phase is the operation phase that involves controlling the real-time system using a digital twin. \par

\subsection{Design Aspects}
\begin{figure}[!t]
	\centering
	\captionsetup{justification=centering}
	\includegraphics[width=8cm, height=5.5cm]{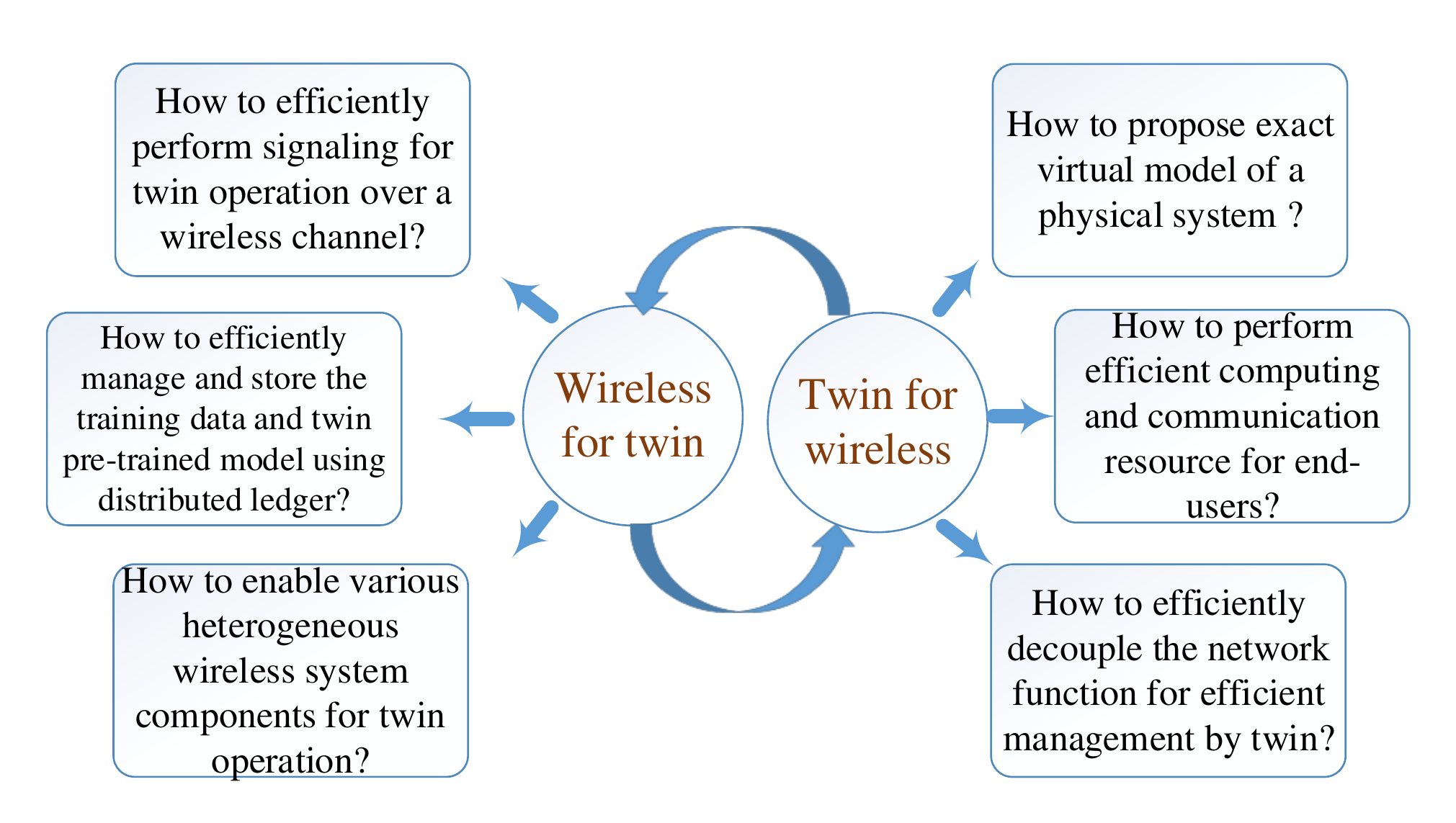}
	\caption{Digital twin and wireless system: design aspects and challenges.}
	\label{fig:aspects}
\end{figure}

In this tutorial, we consider both aspects of digital twins and wireless systems, such as (a) twins for wireless and (b) wireless for twins, as shown in Fig.~\ref{fig:aspects}. The twin for wireless deals with the design of twins to enable various network functions/ applications. On the other hand, wireless for twins deals with efficient communication modeling to enable signaling for twins implementation. Twin for wireless mainly deals with the efficient implementation of wireless system services (e.g., extended reality) using a digital twin. A twins-enabled wireless network will have a variety of players to enable the successful operation of IoE services. These players are blockchain networks, edge/cloud servers, data decoupling interfaces, function decoupling interfaces, and physical devices. Furthermore, the requirements of various IoE applications are significantly diverse. Therefore, special care must be taken while designing twin-based wireless systems to meet the diverse requirements. \par 
Wireless for twins mainly deals with the wireless resource optimization for twinning over wireless networks. Wireless resources can be used in twinning for two operations: (a) twin objects training and (b) twin operation signalling. Twin training will use wireless resources to transfer the data and learning updates. For centralized learning, wireless channel is used to migrate the end-devices data to the shared storage (i.e., servers installed at the edge/cloud. However, some practical scenarios (e.g., autonomous driving cars) generate significant amount of data frequently. Transferring such an enormous amount of data to the shared storage will use significant amount of communication resources. To address this issue, one can use distributed learning (e.g., federated learning) that is based on sending of only learning model update rather than the whole data, and thus consumes fewer communication resources. To train digital twin models using distributed learning over wireless networks, significant amount of resources will be required. Meanwhile, the wireless channel will degrade the performance of distributed learning-based twin model during the transfer of learning model updates between end-devices and aggregation server \cite{chen2020joint}. Additionally, variable latencies will be induced for transferring of learning model updates between devices and twin model servers due to different signal-to-interference-plus noise ratio (SINR). The SINR of devices depends on many factors, such as resource block bandwidth, interference from other users using same resource blocks, and distance between the device and twin model server. For a twin model update of size $u$ and channel throughput $\Gamma$, the transmission latency can be given by $d=\frac{u}{\Gamma}$. The transmission latency can be minimized by many ways, such as (a) reducing the size of twin model updates, (b) enhancing the throughput, and (c) improving the SINR. SINR can be enhanced by optimally performing wireless resource allocation, association of devices with edge/cloud server, and transmit power allocation \cite{khan2020self,han2012game}.

\begin{figure*}[!t]
	\centering
	\captionsetup{justification=centering}
	\includegraphics[width=18cm, height=16cm]{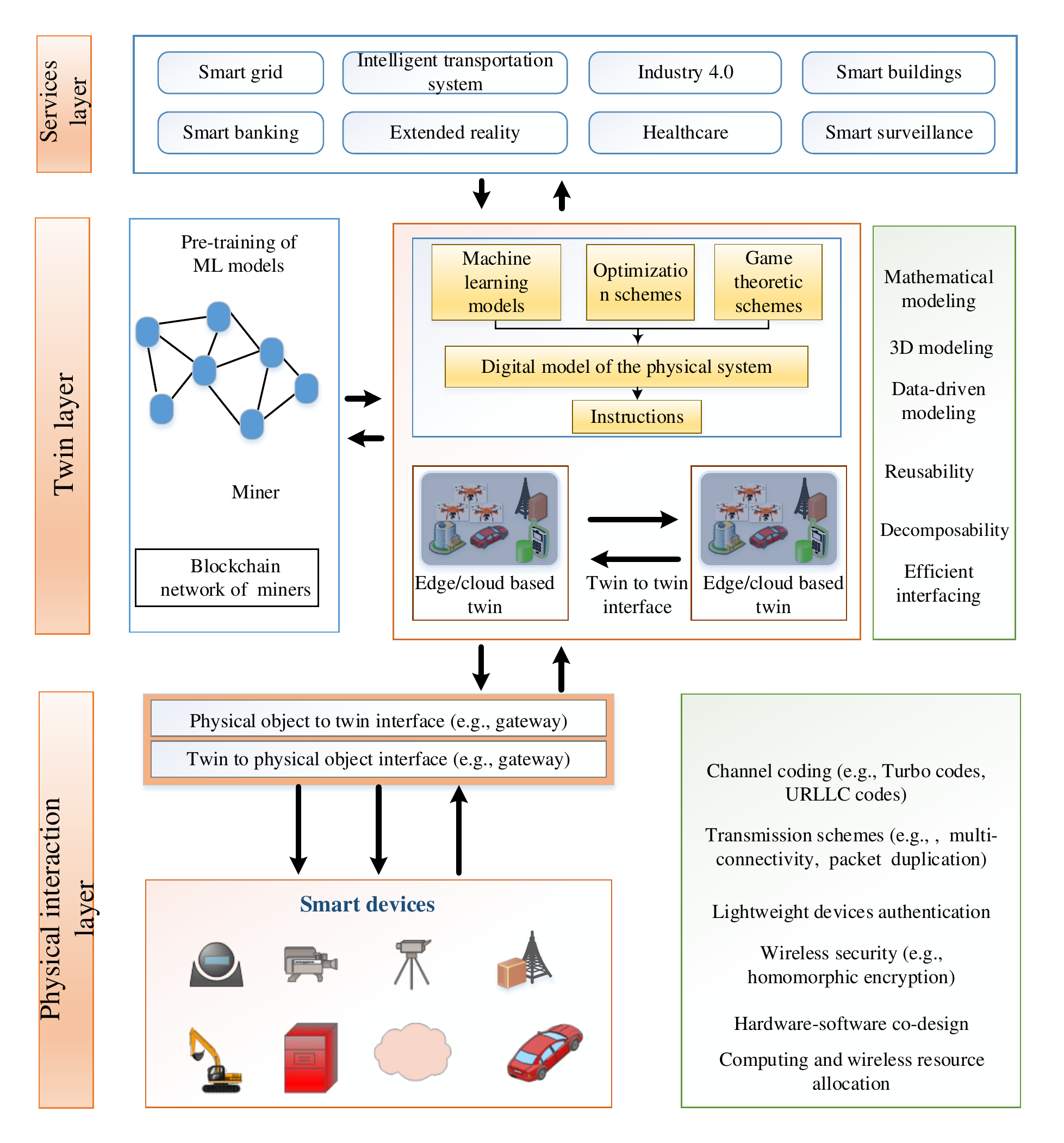}
	\caption{High-level architecture of digital twin enabled wireless systems.}
	\label{fig:architecture}
\end{figure*}

\subsection{High-Level Architecture}
\label{architecture}
A high-level architecture of a digital twin for wireless systems based on logical twin objects is shown in Fig.~\ref{fig:architecture} \cite{khan2021digital,wu2021digital,pires2019digital}. The architecture can be divided into three layers, such as physical devices interaction layer, twin objects layers, and services layer. The services layer contains interfaces for applications. One can request a service from a twin-based wireless system. In response to the user request, semantic reasoning schemes are used to translate the request. Then, the translated requests are passed to the next layers. Next, there is a twin layer that contains logical twin objects. Twin objects use a virtual representation of the physical object/ phenomenon. Representing a physical object/phenomenon using a virtual model faces some challenges. It is difficult to exactly model the physical objects/ phenomenon. One can various ways, such as mathematical model, 3D model, and data-driven model \cite{rasheed2019digital,elayan2021digital}. Representing a physical object using a mathematical model needs several assumptions \cite{ali20206g}. Another way can be to use 3D modeling. However, both mathematical and 3D models may not accurately model the physical model/ phenomenon. To address these limitations, one can use data-driven models-based machine learning to effectively model the physical objects \cite{kumar2019machine, jagannath2019machine, mahdavinejad2018machine,sudharsan2021machine}. The selection of an effective machine learning model is challenging and the right selection of machine learning model consumes a significant amount of time. Note that the twin layer objects can be implemented using container/s or virtual machine/s \cite{khan2020edge,khan2021digital}. Also, one can deploy twin objects either at the network edge or remote cloud. Implementing a twin object at the remote cloud can offer more computing power but at the cost of high latency \cite{fernando2013mobile,coutinho2015elasticity}. To address this issue, one can use a twin object based on the network edge. However, twin objects deployed at the network edge can have low computing power \cite{khan2019edge, abbas2017mobile}. Therefore, one must make a tradeoff between computing power and latency. The last layer is the devices' physical interaction layer. The physical devices layer contains all the physical devices, such as end-devices, edge/cloud servers, base stations, miners, and core network switches, among others. Note that there must effective interfaces between different layers of the twins-based wireless system. These interfaces can be twin to physical object interface, twin to twin interface, and twin to service interface.  \par

Now we discuss the reliability of digital-twin-enabled wireless systems. There are two issues: twin reliability and twin-based service reliability \cite{twin_service_reliability}. Twin reliability refers to the operation of twin without minimum possible interruption due to failure of edge/cloud server running the twin object. An IoT service based on a single twin deployed at the cloud has lower reliability than an IoT service based on multiple twin objects deployed at edge servers. However, the management of multiple twins for a certain service will offer more complexity. Therefore, we must make a tradeoff between the twin reliability and complexity. Additionally, to ensure reliable twin signaling, there is a need to employ channel coding schemes (e.g., URLLC codes) and other techniques (e.g., multi-connectivity, packet duplication). On the other hand, twin service reliability mainly depends on wireless channel reliability and reliable edge/cloud computing. For service wireless channel reliability, similar to twin signaling we can use channel coding schemes and other techniques for communication \cite{lun2008coding,guo2012practical, hausl2009joint}. Additionally, a digital twin can be used for predictive maintenance of $6$G systems to avoid system malfunctions and cyber-attacks through artificial intelligence analytics and simulation. The summary of the steps for twin-based wireless system operation is as follows. \begin{itemize}
    \item First of all, end-user will request a service. This request will be translated using semantic reasoning schemes to make it compatible with the twin object-based architecture.
    \item Next, the twin object will be created to to serve the end-user.
    \item Finally, the twin object uses mathematical optimization and machine learning schemes in addition to virtual representation to enable efficient resource optimization for various services.
\end{itemize}

\subsection{Frameworks}
In this section, we discuss various frameworks designed for implementing digital twins. Moreover, we critically analyze them and discuss their advantages and limitations.\par
\subsubsection{Eclipse Ditto}
Eclipse Ditto is an open-source framework for creating digital IoT twins \cite{twin_frameowrk2}. The framework consists of Ditto services (components), external dependencies (MongoDB and nginx), and application programming interfaces (APIs). Ditto consists of microservices, each with its own data store where reading and writing take place. Every microservice has a set of APIs (events, command responses, commands). Moreover, a microservice can communicate with other microservices using defined signals. Within a Ditto cluster, all microservices can communicate asynchronously using an open-source toolkit, namely, AKKA remoting. Therefore, each service can acts client, enabling a TCP endpoint, and a server. The messages between microservices are serializable (i.e., from Java objects to JavaScript Object Notation(JSON)) and deserializable (i.e., from JSON to Java objects).


\subsubsection{Model Conductor-eXtended Framework}
Model Conductor-eXtended (MCX) is an open-source framework for digital twins experimentation \cite{MCXtwin_frameowrk,MCXtwin_frameowrk2}. The framework enables us to co-execute the digital systems and physical systems as well as asynchronous communication. Furthermore, support for machine learning models, customized models, and running FMUs (simulation models packaged according to the FMI specification) is also provided. The MCX uses standard data transmission protocol, time-synchronous implementation of time-consuming simulation models, and decoupling of the queue and the model computation module, for offering scalability which is one of the important design aspects of IoT applications.


\subsubsection{Mago3D}
Mago3D is an open-source digital twin platform developed by a South Korean company, namely, Gaia3D inc. \cite{Mago3D_frameowrk1}. The purpose is to real-world objects, phenomenon, and process on web environment \cite{Mago3D_frameowrk2}. The platform consists of a geospatial data server, data conversion server, platform core server, and web server, for realizing the digital twin-based architecture. Mago3D has been applied in various sectors, such as Korean national defense, indoor data management, shipbuilding, and urban management, and have shown good results.\par 
All of the aforementioned digital twin platforms (i.e., Eclipse Ditto,Model Conductor-eXtended Framework, and Mago3D) are promising towards the realization of digital twin-based systems. However, none of them effectively considered the effects of wireless channels on the performance of digital twins. In wireless systems, wireless channel uncertainties will significantly affect the performance of IoT applications. Therefore, one must effectively model these wireless channel uncertainties while digital twinning for wireless system applications. \par  


\subsection{Summary and Lessons Learned}
In this section, we presented the fundamentals of digital twins and derived the definition of twins in the context of wireless systems. We also presented the key design aspects, such as twins for wireless and wireless for twins. Several available frameworks for digital twins experimentation are also discussed. Some key lessons learned here include:
\begin{itemize}
    \item A digital twin should be designed in a generalized way so that it can be easily reusable for future services. Designing a twin takes significant effort and time. Twins based on machine learning should be trained using more data to make them generalized so as to use them for multiple scenarios. Such kinds of twins will minimize the design of the service based on twins. Furthermore, this twin will reduce the design cost.
    \item Mostly, the current digital twin frameworks are designed without effectively taking into account the wireless channel impairments. There is a need to propose a novel framework for digital twinning over wireless networks. The digital twin framework should consist of multiple base stations/access points and devices. The wireless impairments degrade the SINR of the transmitted signal. The SINR can be improved using effective resource allocation, association, and transmit power allocation. Therefore, the framework allows us to enable effective resource allocation, association, and power allocation.
    \item Digital twins bring together two main use cases for wireless systems: twin for wireless and wireless for twins. Twin for wireless is necessary for effective use of resources (e.g., computing, communication, transmit power). On the other hand, wireless for twins deal with resource-efficient management for twins signaling. Therefore, it is necessary to effectively manage both aspects of digital twin and wireless systems.    
\end{itemize}

\begin{figure*}[!t]
	\centering
	\captionsetup{justification=centering}
	\includegraphics[width=18cm, height=22cm]{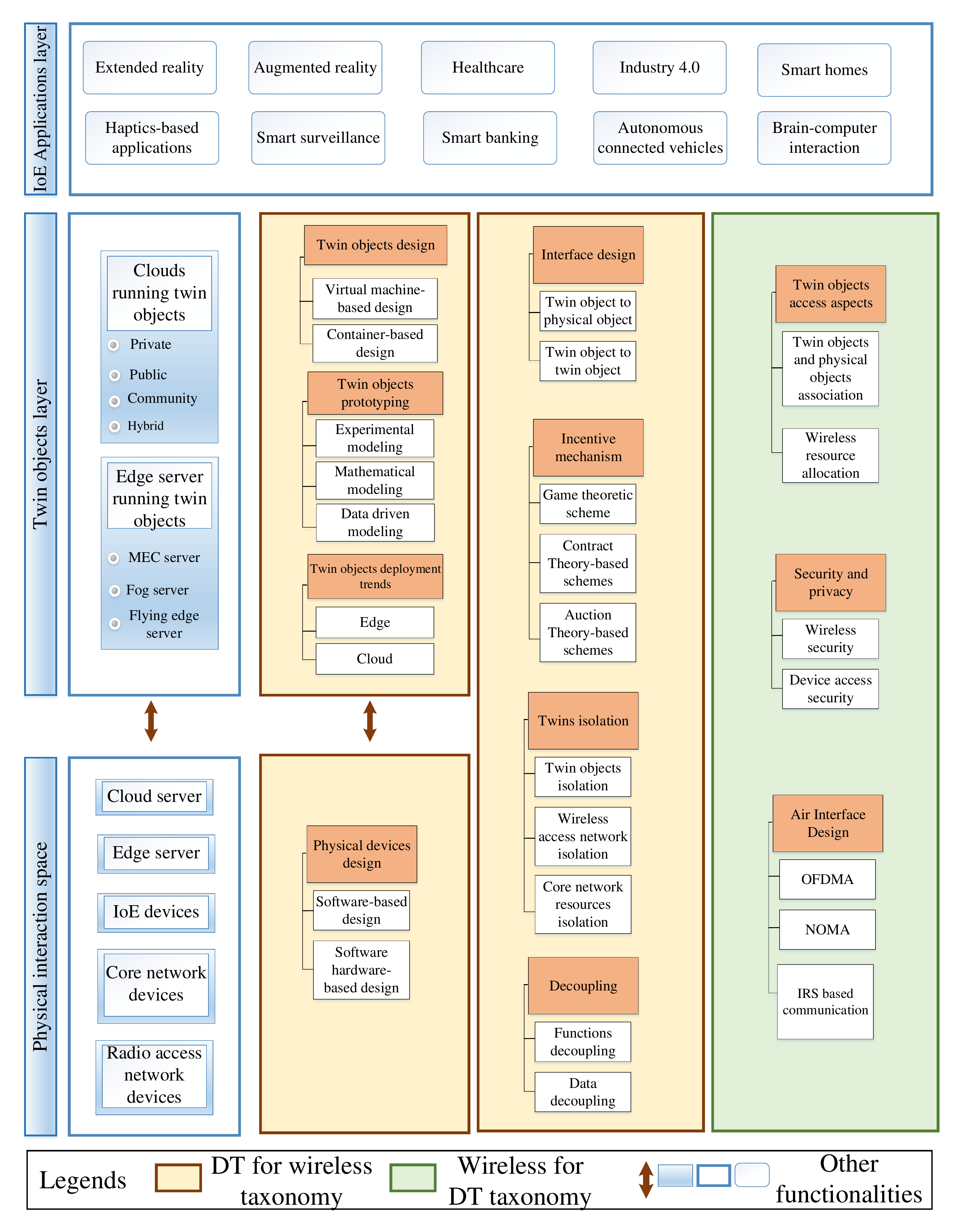}
	\caption{Taxonomy of digital twin and wireless systems.}
	\label{fig:taxonomy}
\end{figure*}


\section{Taxonomy: Twins for Wireless}
Both aspects (i.e., wireless for twins and twin for wireless) of twin-based wireless system deal with a variety of players that act in a complex environment to enable digital twin-based wireless systems. To classify the areas for study in both aspects, we devise a taxonomy that covers both \textit{twins for wireless} and \textit{wireless for twins}, as shown in Fig.~\ref{fig:taxonomy}. Fig.~\ref{fig:taxonomy} divides the taxonomy into layers, such as physical interaction later and twin later. The parameters are positioned according to their role at the layer. For instance, \textit{physical devices design} is positioned at the physical interaction layer due to its role there. Additionally, the taxonomy related to twins for wireless and wireless for twins is shown by different colors in Fig.~\ref{fig:taxonomy}. Twin for wireless refers to the enabling of various wireless systems applications, such as extended reality, human-computer interaction, and digital healthcare, using twins. We categorize the \textit{twins for wireless} aspect research into twin isolation, incentive design, twin object design, twin object prototyping, twin objects deployment trends, physical end-devices design, decoupling, and interfaces design. Twin isolation deals with the seamless operation of twin-based services without interrupting the performance of other twins. Incentive design enables attractive incentives for devices that participate in twinning. Twin objects design helps in the on-demand creation of twins using existing computing hardware for various applications, whereas twin objects prototyping helps in the creation of virtual models of the wireless systems. Twin object deployment trends guide us about the placement (e.g., cloud or edge) of twin objects. Decoupling allows the seamless operation of the twin-based services with minimum possible dependency on the underlying hardware. On the other hand, interfaces are used for various types of communications, such as among twin objects, and between twin objects and physical devices.   \par   

\subsection{Twins Isolation}
Twins isolation deals with the operation of a twin-based application (e.g., extended reality) without affecting the performance of other twin-based applications (e.g., brain-computer interaction). The operation of various twins for different applications requires computing, storage, and communication resources. The computing resources are used for performing blockchain mining as mentioned in architecture (for more details, please refer to Fig.~\ref{fig:architecture}) in Section~\ref{architecture}. Meanwhile, there is a need for storage to store pre-trained twin models for various applications. Additionally, sufficient communication resources are required for the signaling of twin instructions. To do so, there can be mainly two ways of enabling twins, such as using dedicated hardware and shared hardware. Using dedicated hardware for twin-based applications can better perform but at the cost of high expenditure. Therefore, using dedicated hardware for the end-to-end implementation of twin-based applications is not feasible. To address these issues, one can use shared hardware for twin-based applications. However, sharing hardware for various twin-based applications may affect performance. Therefore, there is a need for isolated operation of twin-based applications. \par           
\begin{table*}

\caption {Virtualization schemes for radio access networks.} \label{tab:miner comparison} 
  \centering
  \begin{tabular}{p{2.2cm}p{4cm}p{4cm}p{4cm}}
    \toprule 

     \textbf{Reference} & \textbf{Network details} & \textbf{Utility} & \textbf{Solution approach}   \\
     \midrule
    Ludwig \textit{et al.}~\cite{ludwig2020efficient} & A graph of nodes (i.e., cloud nodes and end-users nodes) as vertices with connectivity cost (e.g., latency) as vertices was considered.. & Acceptance ratio that is given by dividing embedded (i.e., served) slices divided by total slices. & Heuristic approach \\
     \midrule
     De Bast~\textit{et al.}~\cite{de2019deep} & A set of Wi-Fi users with different, dynamic requirements was considered.  & A reward function for optimizing throughput was used. & Double deep Q-network-based scheme. \\
    \midrule 
   Kim~\textit{et al.}~\cite{kazmi2020distributedss}  & A set of infrastructure provides, users, and set of virtual network operators.  & To jointly maximize the network throughput and profit of the revenue of the infrastructure provides. & Match theory-based solution.  \\
    \midrule 
  Manh ho~\textit{et al.}~\cite{ho2018network}   & A heterogeneous cellular network providers, users, and virtual network operators was considered. & To optimize overall energy and revenue of the virtual network operators. & Main problem is decomposed into two sub-problems that are solved using two-stage Stackelberg game, matching theory, and iterative schemes, respectively.  \\  
   \midrule
    Kazmi~\textit{et al.}~\cite{kazmi2017matching}  & A set of virtual network operators, set of infrastructure providers, and users were considered.   & To jointly maximize the utilities of infrastructure providers (i.e., profit) and virtual network providers (i.e., maximize the end-users throughput and minimize the cost).  & Matching theory-based solution \\

\bottomrule 
\end{tabular}
\end{table*}

Isolation in digital twin-enabled wireless systems can be divided into twin objects isolation, core network isolation, and access network isolation \cite{kazmi2019network,khan2020network}. Radio access network isolation refers to using the same access network for various twins without affecting the performance of other twins. Access network isolation can be performed effectively with better management using radio resource virtualization \cite{chang2018radio, datsika2016matching, kazmi2017hierarchical}. In resource virtualization, the third party, a mobile virtual network operator (MVNO) will buy radio resources from multiple network operators. For twins, using resources from different network operators is challenging and suffers from high management complexity. Therefore, one can easily use network virtualization that will enable MVNO to buy network operators' resources and then sell the resources (i.e., resource partitioning) to various twins based on their requirements in such a way to fulfill the end users' QoS. A twin can also buy resources from multiple MVNOs. In the virtualization approach, the twin does not require management of the resources of multiple network operators, and thus avoids high management complexity. However, this will increase the management complexity of the overall system. Numerous approaches that perform resource selling between the MVNOs and twins are heuristic schemes, deep reinforcement learning schemes, and game-theoretic schemes. Heuristic schemes will offer high implementation complexity \cite{ludwig2020efficient}. Deep reinforcement learning schemes many also offer high training complexity \cite{de2019deep}. Therefore, one can preferably use matching game-based schemes for resource selling between the MVNOs and twins \cite{kazmi2020distributedss,ho2018network,kazmi2017matching}. \par
Similar to radio access network isolation, one can perform core network isolation. However, the core network isolation will be different than the access network isolation. The core network can be constituted using various technologies, such as optical fiber, microwaves, and wired communication technologies. Therefore, performing resource partitioning at the core network for twins may be more challenging. There is a need to propose a hybrid scheme for core network isolation that has different criteria for each communication technology (e.g., optical fiber links, microwave links). On the other hand, twin objects isolation can be performed using shared computing hardware (e.g., cloud) for multiple virtual machines running twin objects. Isolation of such twin objects is necessary for improving performance and avoiding security attacks \cite{Twin_iso1,Twin_iso2}. For instance, if the twin objects are not well isolated, then the security attack instantiated within a twin object can easily affect other twin objects. One easy way is to assign dedicated hardware for running a twin object. However, this approach is not feasible due to the high implementation cost. Therefore, there is a need to share the same hardware for virtual machines running twin objects. Similar to radio access network resource sharing, one can use various schemes based on heuristic algorithms, matching game, and deep reinforcement learning.\par

\begin{figure}[!t]
	\centering
	\captionsetup{justification=centering}
	\includegraphics[width=8cm, height=7cm]{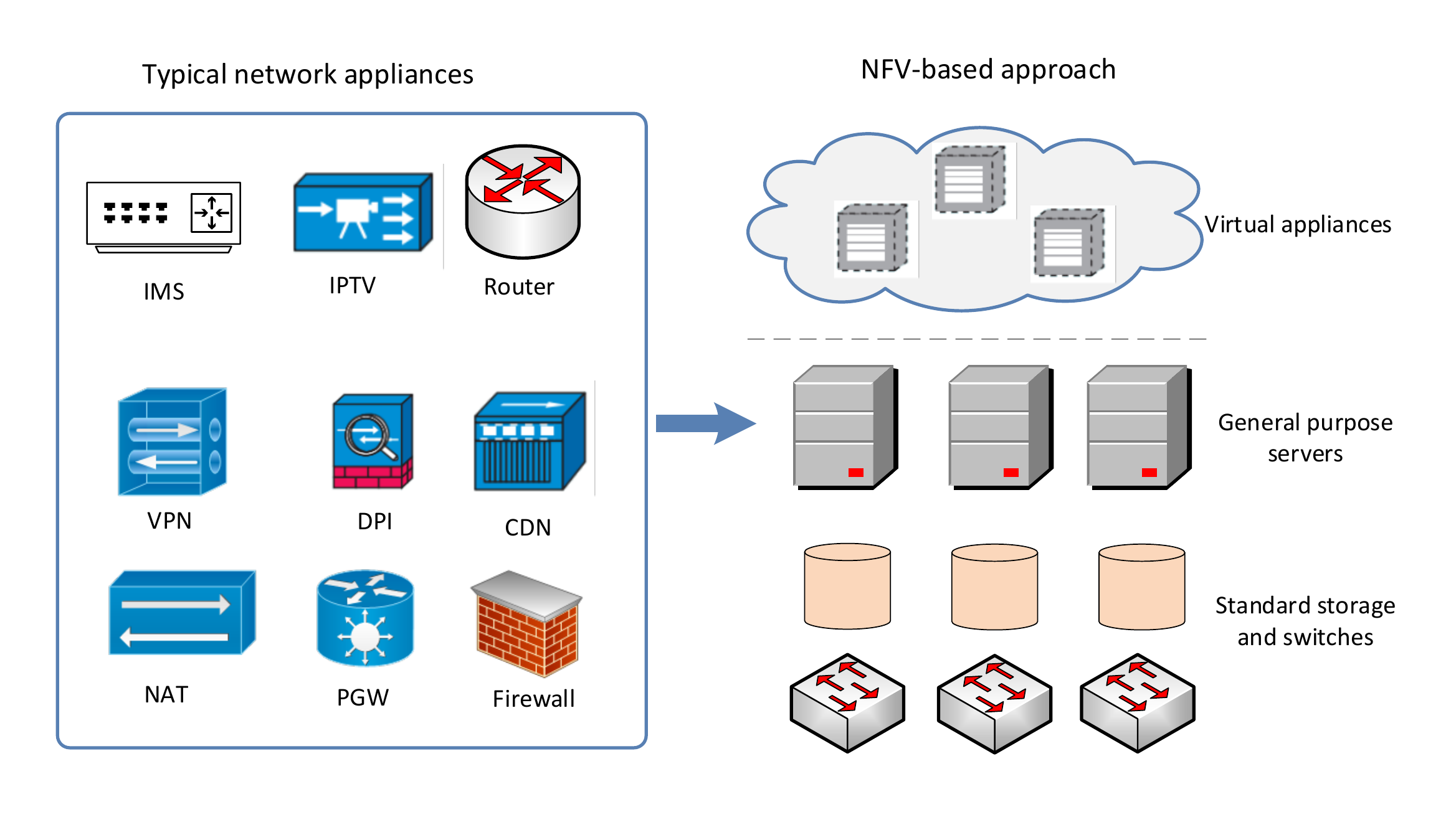}
	\caption{Overview of network function virtualization.}
	\label{fig:NFV}
\end{figure}

\begin{figure*}[!t]
	\centering
	\captionsetup{justification=centering}
	\includegraphics[width=18cm, height=8cm]{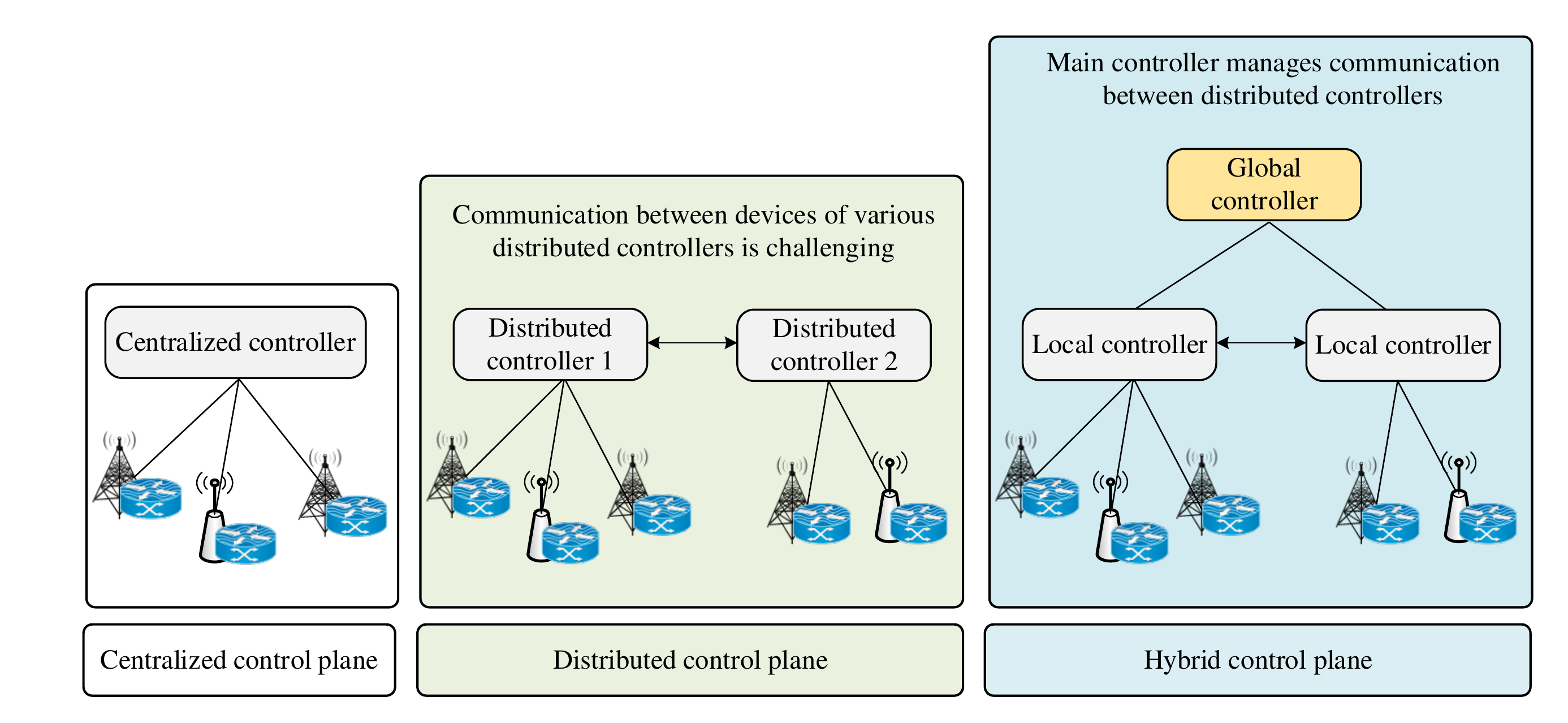}
	\caption{Twins function decoupling using (a) centralized controller, (b) distributed controllers, and (c) hybrid controllers.}
	\label{fig:sdnplanes}
\end{figure*}

\subsection{Decoupling}
Decoupling in digital twin-enabled wireless systems can be of two types, such as functions decoupling and data decoupling \cite{khan2021digital}. Data decoupling can be performed using data homogenization techniques. This will enable the twin-based system design independent of the underlying the smart devices network. The layered architecture of twin-based wireless system is discussed in Section~\ref{architecture}. Fig.~\ref{fig:architecture} shows that there will be a variety of smart devices (e.g., temperature sensors, image sensors) at the physical interaction layers. To transform these data into homogenized data, there is a need to employ efficient data homogenization techniques. On the other hand, functions decoupling enables the twin-based system to decouple the management functions from the physical interaction layer to the twin layer. Function decoupling can be implemented by network slicing that is based on NFV and SDN \cite{bera2017software,mohammed2020review,kalkan2017securing,han2015network,mijumbi2015network,joshi2016network}. Network slicing offers the use of shared network resources for various applications (e.g., intelligent transportation system, healthcare). Network slicing uses NFV and SDN to separate the control plane from the physical interaction plane for easier and efficient management. NFV is based on implementing various network functions using generic hardware as shown in Fig.~\ref{fig:NFV}. Implementing various network functions using generic hardware will offer us a cost-efficient design. To implement the physical devices using the NFV-based approach, there may few limitations, such as throughput and latency deterioration \cite{han2015network}. The per-instance capacity of the NFV-based device may not be similar to the actual physical device. There is a need to keep such deterioration in performance lowest possible. Other challenges faced by NFV-based design are reliability and security issues. Multiple NFVs for various network functions running on the same hardware might result in failure due to a physical damage/ security attack \cite{lal2017nfv}. To address this issue, one can migrate the NFV to a new device. However, migration may introduce a delay that is undesirable in many IoE/IoT applications \cite{wang2018survey,hassan2019edge,hu2015mobile}. Additionally, mobility of the end-devices also requires the migration of NFV-based functions to the new network edge. To enable such migration, we can use machine learning-based mobility prediction schemes to proactively determine the next location \cite{khan2020edge}. On the other hand, NFV-based implementation faces new security concerns. NFV-based functions running on remote cloud that is not controlled by the network operators. Therefore, security must be ensured for the NFV-based functions running on the cloud. On the other hand, SDN separates the control plane from the data plane. In a typical SDN control plane, a single centralized controller is used to control multiple switches for performing various functions (e.g., routing) \cite{karakus2017survey,amiri2019efficient,sarmiento2021decentralized,bannour2017distributed}. These switches intended to control wireless system devices can be installed on dedicated hardware for better performance. However, this approach will increase the implementation cost of the system. To address limitations, we can design SDN-based devices with built-in capabilities of SDN switches \cite{bannour2017distributed}. Controlling SDN-based devices using a single centralized controller may suffer from scalability and reliability issues which are one of the most important design considerations of wireless systems. An increase in the number of devices will increase the signaling between SDN-based devices and the centralized controller, and thus suffer from high latency issues. To address this limitation, one can use a distributed control plane that consists of multiple SDN controllers \cite{togou2019hierarchical,ahmad2021scalability}. Every controller among a set of controllers controls some end-devices, and thus minimizes the latency due to signaling. However, using multiple controllers will lead to an increase in control plane management complexity. Additionally, communication between the devices connected to different distributed SDN controllers may suffer from extra latency in communication. To reap the advantages of both centralized control and distributed control planes, one can use a hybrid control plan. A hybrid control plane will use both centralized and distributed controllers \cite{huang2016hybridflow,fu2015hybrid}. In the hybrid control plane, the main controller controls the local controllers. The local controllers control its small set of devices. All the functions that are local to the devices are handled by the local controller, whereas the functions that require global network knowledge (e.g., wireless resource allocation) are controlled by the main controller in addition to local controllers. Overview of the centralized, distributed, and hybrid control planes are given in Fig.~\ref{fig:sdnplanes}. Meanwhile, the tabular comparison is given in Table.~\ref{tab:snd comparison}. The management complexity of the centralized control plane is less than the distributed control plane and hybrid control plane. Additionally, the signaling latency of the centralized control plane will be higher than the distributed and hybrid control plane.

\begin{table}

\caption {Comparison of twins function decoupling.} \label{tab:snd comparison} 
  \centering
  \begin{tabular}{p{2.2cm}p{1.5cm}p{1.5cm}p{1.5cm}}
    \toprule 
   
     \textbf{Twin function type} &  \textbf{Management complexity} & \textbf{Latency} & \textbf{Implementation cost}  \\
     \midrule
    Centralized controller-based function decoupling & Lowest & High & Lowest \\
    \midrule 
    Distributed controllers-based function decoupling & High & Low & High \\
    \midrule 
    Hybrid controller-based function decoupling & Highest & Low &  Highest \\

\bottomrule 
\end{tabular}
\end{table}

\subsection{Interfaces Design}
Interface in digital twin-based system can be many types, such as user to twin system interface, twin to object interface, twin object to twin object interface, and air interface. Detailed discussions about air interface will be given in Section~\ref{air interface}. Here, we discuss twin to physical objects interface, user to twin system interface, and twin object to twin object interface. The twin to physical object interface will can be both wireless (e.g., industrial machine end-devices) and wired (e.g., smart phone running an application). Similarly, twin to twin interface can also be wired and wireless. For distributed twins deployed at multiple edge servers \cite{khan2021digital}, there is a need to enable wireless communication among them. Another way of communication can be through the core network \cite{khan2020edge}. On the other hand, there is a need to propose efficient and easy to use interface for communication between end-users and the system. The interface can be of various types, such as voice interface (e.g., Amazon Echo), touch interface (e.g., smart phone), and physical button interface \cite{interface_2}. These interfaces enables users to interact with the twin-based system. For instance, smart phones uses iOS, Samsung One UI, OxygenOS, Android One, and Indus OS, are the operating systems used by Apple, Samsung, OnePlus, Google’s Android One programme, and India, respectively \cite{interface_1}. For smart phone, one can use any of the aforementioned interfaces of operating systems. However, for applications based on digital twins, there may be a need to propose new interfaces because of their different architecture than the existing ones.\par

\subsection{Twin Objects Design}
Twin objects are instantiated upon request from the end-users. There are two ways to instantiate the twin objects, such as virtual machine-based twin objects and container-based twin objects \cite{khan2019edge,khan2020edge}. A virtual machine can be defined as the architecture that is independent of the underlying hardware. Virtual machines can be mainly of two types, such as system virtual machine and process virtual machine \cite{virtual_machine_1}. In the context of twin-based wireless system, the system virtual machine can be used to model a complete IoT service (e.g., AR-based healthcare system), whereas the process virtual machine can be used to model the particular portion (e.g., edge caching module for smart infotainment system) of digital twin-based system. Modeling of a complete system may be easier than modeling of a particular part. The reason for this can be decoupling only a part of a system from the hardware is challenging. Note that the virtual machine is different from the operating system. In an operating system, language independent extensions of hardware are created, whereas a virtual machine creates a machine-independent instruction set. Note that virtual machine-based virtualization can be seen as an infrastructure-as-a-service (IaaS) component that can enable the same hardware via virtualization for running multiple operating systems (e.g., twins). This operational approach has a main drawback of high management complexity. To address this issue, one can use container-based twin objects. Containers can help minimize management complexity by running multiple twins on the same operating system \cite{vM_1}. In fact, containers can be considered as operating systems-as-a-service. Therefore, one can easily implement twin objects using container based design due to its low management complexity and light weight nature compared to virtual machine based design \cite{vM_2}. However, container-based twin objects may face more security attacks than the virtual machine-based design. The reason for this is the same operating system running various twin objects, and thus more prone to security attacks. On the other hand, virtual machine based twin objects has less security attacks due to their IaaS nature (i.e., multiple operating systems are used to implement twins on the same hardware). Virtual machine-based twins achieve this at the cost of high management complexity. Therefore, a tradeoff must be made between the management complexity and security.

\subsection{Twin Objects Prototyping}
True prototyping of twin objects is one of the primary challenges in digital twinning over wireless networks. Physical objects are characterized by a set of attributes (e.g., shape, mass, energy). These attributes are difficult to exactly model \cite{Akhan2021survey}. Additionally, these parameters will significantly affect the performance of the system. Therefore, one must effectively model these parameters in digital twinning. In experimental modeling, a series of experiments are performed to find out the various parameters of a physical wireless system \cite{tang2020wireless}. Various entities of a wireless system can be modeled using experimental data. For instance, the work in \cite{tang2020wireless} modeled the free space path loss for intelligent reflecting surfaces using a series of experiments. Their experimental setup was comprised of accessories (i.e., cables and blocking object (electromagnetic wave-absorbing materials), Rx horn antenna, RF signal analyzer (Agilent N9010A), Tx horn antenna, RF signal generator (Agilent E8267D), and metasurface. Although experimental modeling of a wireless system can provide many benefits, it has a few issues \cite{tang2020wireless}. These issues are high costs associated with the experimental setup and human experimental errors. Additionally, if want to design new service twins, there is a need to perform many experiments that will significantly increase the cost, and thus may be undesirable \cite{khan2021digital}. To address this issue, one can use mathematical modeling that is based on a mathematical representation of the wireless phenomenon. Although mathematical models can be obtained in a cost-effective way, there is a need to pay special attention while modeling various parts. Typical mathematical models are based on various assumptions, therefore, we must make assumptions that are close to real time scenario for more practical results.\par 
Mathematical models can be used to model a variety of scenarios, there are some scenarios where it seems difficult to model using mathematical models \cite{ali20206g}. To address this limitation, one can use data-driven modeling. All the wireless system applications (e.g., XR) generate a significant amount of data that can be used in modeling their behavior \cite{khan2020edge}. One can use the data of wireless system applications to train machine learning models. Note that machine learning models can model the wireless system phenomenon that can not be modeled using mathematical techniques. However, such training will be at the cost of the training cost. There can be mainly two possible ways to train machine learning models, such as centralized machine learning and distributed machine learning. Centralized machine learning relies on centralized training at the remote cloud or edge server. Although centralized machine learning offers many advantages, it has a prominent issue of privacy leakage because of transferring end-devices data to the centralized cloud/edge server. To address this issue, one can use distributed machine learning. In distributed machine learning, end-devices iteratively train their local model using the local datasets. The local models are then sent to the edge/cloud server for aggregation to yield a global model. Although distributed machine learning can better preserve users' privacy, it has its own challenges, such as data heterogeneity, system heterogeneity, and wireless channel uncertainties.\par

\begin{figure*}[!t]
	\centering
	\captionsetup{justification=centering}
	\includegraphics[width=18cm, height=14cm]{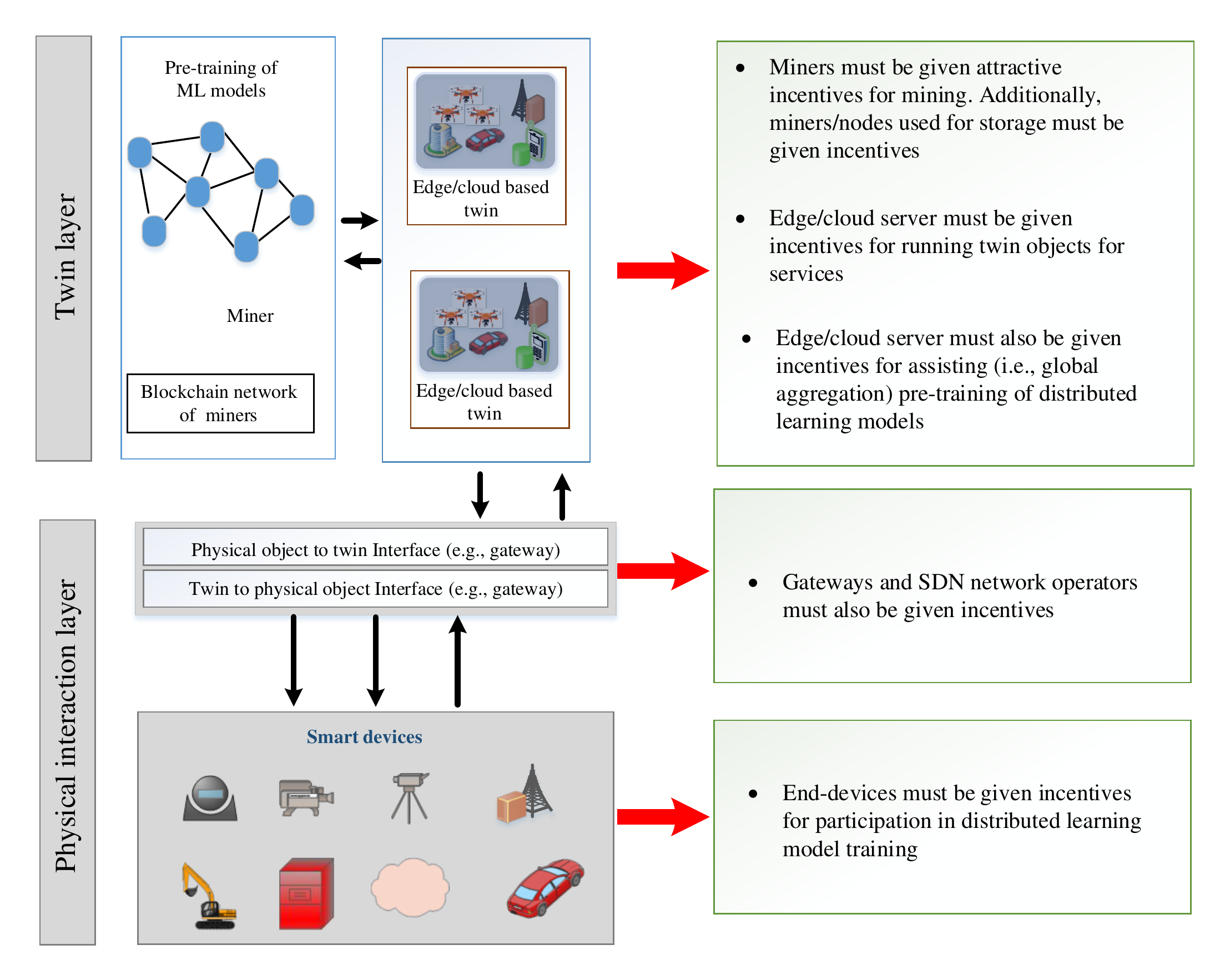}
	\caption{Overview of incentives in digital twin-based wireless systems.}
	\label{fig:incentive}
\end{figure*}

\begin{table*}
 
\caption {Federated learning incentive mechanisms \cite{khan2021federated}.} \label{tab:incentive mechanism} 
  \centering
  \begin{tabular}{p{2.5cm}p{2.5cm}p{3cm}p{3cm}p{3.5cm}}
    \toprule 
   
     \textbf{Reference} &  \textbf{Incentive model} & \textbf{Motivation} & \textbf{Device utility/ bid} & \textbf{Utility of aggregation server} \\
     \midrule
    
   Pandey \textit{et al.},~\cite{pandey2019crowdsourcing} & Stackelberg game &  To enable accurate, communication efficient federated learning model. & Difference between cost (i.e., communication and computation cost) and reward. & Concave function of $\log(1/global~accuracy)$ \\
    
    \midrule
    Le \textit{et al.},~\cite{le2020auction} & Auction theory  & To reduce the social cost (i.e., devices bids cost). & Difference between cost and reward. & To reduce the bids cost. \\
    \midrule
    
    Kang \textit{et al.},~\cite{8832210} & Contract theory & Motivating reliable end-devices for participating in learning process. &  Difference between energy consumption and reward. & To maximize the profit (i.e., difference between reward given to end-devices and total time for global iteration). \\
    
\bottomrule 
\end{tabular}

\end{table*}

\subsection{Incentive Mechanism}
The nature of the twin-based wireless systems will be different compared to the existing wireless systems. Therefore, the design of incentive mechanisms for the digital twin-based system will be different. Here, we consider the high-level architecture of digital twin-enabled wireless systems (Fig.~\ref{fig:architecture}) presented in Section~\ref{architecture}. The different types of incentives required in digital twin-enabled wireless systems are presented in Fig.~\ref{fig:incentive}. As described in Section~\ref{architecture}, digital twin-based wireless system can use distributed learning to train twin models. These pre-trained twin models are stored using blockchain network in an immutable, transparent manner and can be used in future to serve the requesting users. Upon request from end-users', the twin-based system serves the end-users' by controlling the network entities for efficient management of resources. To do so, there must be incentives to various entities of the digital twin-based wireless system. Incentive mechanisms for the digital twin-based wireless system have three main aspects, such as \textit{incentive mechanism design for pre-training of twin models}, \textit{incentive mechanism design for blockchain mining}, and \textit{incentive mechanism design for control of physical objects}, as shown in Fig.~\ref{fig:incentive}. During these three phases, the players involved are end-devices, aggregation servers, edge/cloud servers running twin objects, and blockchain miners used for storing pre-trained models and training data. Attractive incentives must be given to all the players of a digital twin-based wireless system for performing their tasks. \par
When using distributed learning for digital twins, end-devices participate in the learning model local models \cite{khan2021dispersed}. Moreover, edge/cloud servers run twin objects and also assist (i.e., aggregation of local models) distributed learning. Therefore, there is a need to give them reward for performing their jobs. However, the nature of the incentive mechanism will be different for end-devices and edge/cloud servers for two tasks (i.e., distributed learning and twinning control). For distributed learning-based twins, the edge/cloud servers will interact with the end-devices for learning a global model. End-devices perform local model learning and edge/cloud server performs aggregation of the end-devices local models to yield a global model. The global model is then shared with the end-devices to update their local models. Such an iterative process takes place until convergence is achieved. One can design an incentive mechanism for such a scenario by defining a utility at the edge/cloud server. The utility of edge/cloud servers can be to maximize the global model accuracy. Meanwhile, the devices with high local accuracy can be given more monetary reward. This seems practical approach because generally, the end-devices need to consume more computing resources to obtain a better local model accuracy. Various works proposed incentive mechanism design for distributed learning using Stackelberg game, auction theory, and contract theory \cite{khan2021federated}. An overview of these incentive mechanisms is given in Table~\ref{tab:incentive mechanism}. Additionally, details about the incentive model approach, motivation, device utilities, and aggregation server utilities are given.\par

On the other hand, for a twinning operation, incentives must be provided to the edge/cloud servers running twin objects. In digital twin-based architecture, pre-trained twin models must be stored using blockchain for future use \cite{khan2021digital}. Miners used for performing blockchain consensus algorithms and storage must be given attractive incentives \cite{ommunication}. In \cite{ommunication}, a Fee and Waiting Tax scheme was proposed to provide miners with an incentives for mining and storage of the generated blockchain blocks. The Fee and Waiting Tax scheme was based on addressing the two kinds of issues associated with the existing blockchain protocols, such as selecting a transaction by a certain miner will impose storage cost on the other miners and users' transactions generation may increase the waiting time for other users. The Fee and Waiting Tax has two main steps, such as fee choices and waiting tax. In fee choices, the proposed scheme offers a set of fee-per-byte choices for users while ensuring the miners to get sufficient fees for their work. The second step is that the Fee and Waiting Tax scheme is waiting tax on users proportional to their negative impact on other users. This enables the users to become more conservative while generating transactions. On the other hand, the edge/cloud servers running the twin objects must be given incentives for their contributions. Meanwhile, the SDN network operators must also be given attractive incentives for enabling efficient twinning.  \par

\subsection{Twins Deployment Approaches}
\label{trends}
Twins can be mainly deployed either at the network edge or cloud depending on the requirement of the application. Every application has distinct requirements in terms of latency, quality of physical experience, computing resource requirement, and reliability, among others. Depending on the these requirements one can deploy twins at various locations (e.g., edge). Twins deployed at the network edge can enable services with low latency compared to twins at the cloud. Meanwhile, the twins at the edge can have more context awareness (e.g., end-devices location, mobility-awareness). Although edge-based twins can enable various applications with a variety of advantages, they have limitations in terms of low computing resources. Edge has lower computing resources than the cloud. Therefore, one can deploy twins at the cloud. Cloud has more computing resources but at the cost of high latency and low context-awareness. \par

\begin{table*}
\caption {Comparison of various twin objects deployment \cite{khan2021digital}.} 
\label{tab:edgecomparison} 
  \centering
  \begin{tabular}{p{1.4cm}p{10.5cm}p{1.8cm}p{1.3cm}p{1.3cm}}
    \toprule
   &\textbf{Description}&\textbf{Edge-based twin object} & \textbf{Cloud-based twin object} & \textbf{Edge-cloud-based twin objects} \\
\hline
\textbf{Twins Robustness (reliability)} & It refers to seamless operation during failure of one of twin objects. & Highest (for multiple edge-based twins) & Lowest & Medium\\
\midrule 
\textbf{Geo-distribution} & This metric enable us with the information about the geographical distribution of twin objects. & Distributed & Centralized & Hybrid \\\midrule
\textbf{Elasticity} & This metric shows the ability of a twin-based wireless system in providing highly dynamic requirements of various services via elastic resource allocation. & High & Low & High \\ \midrule

\textbf{Mobility support} & It refers to the ability of a twin-based wireless system to serve the mobile devices seamlessly. & High & Low & Medium \\ \midrule
\textbf{Latency} & It deals with the overall delay in providing the service to end-users. & Low & High & Medium \\ \midrule
\textbf{Context-awareness} & It deals with the ability of a twin-based wireless system to obtain network and devices information. & High & Low & Medium \\\midrule

\textbf{Scalability} & It deals with fulfilling the requirements of a massive number of wireless devices. Additionally, the addition of more devices should not degrade the latency performance. & High & Lowest & Low \\ 
\bottomrule
  \end{tabular}
\end{table*}\par

\begin{figure}[!t]
	\centering
	\captionsetup{justification=centering}
	\includegraphics[width=8cm, height=6cm]{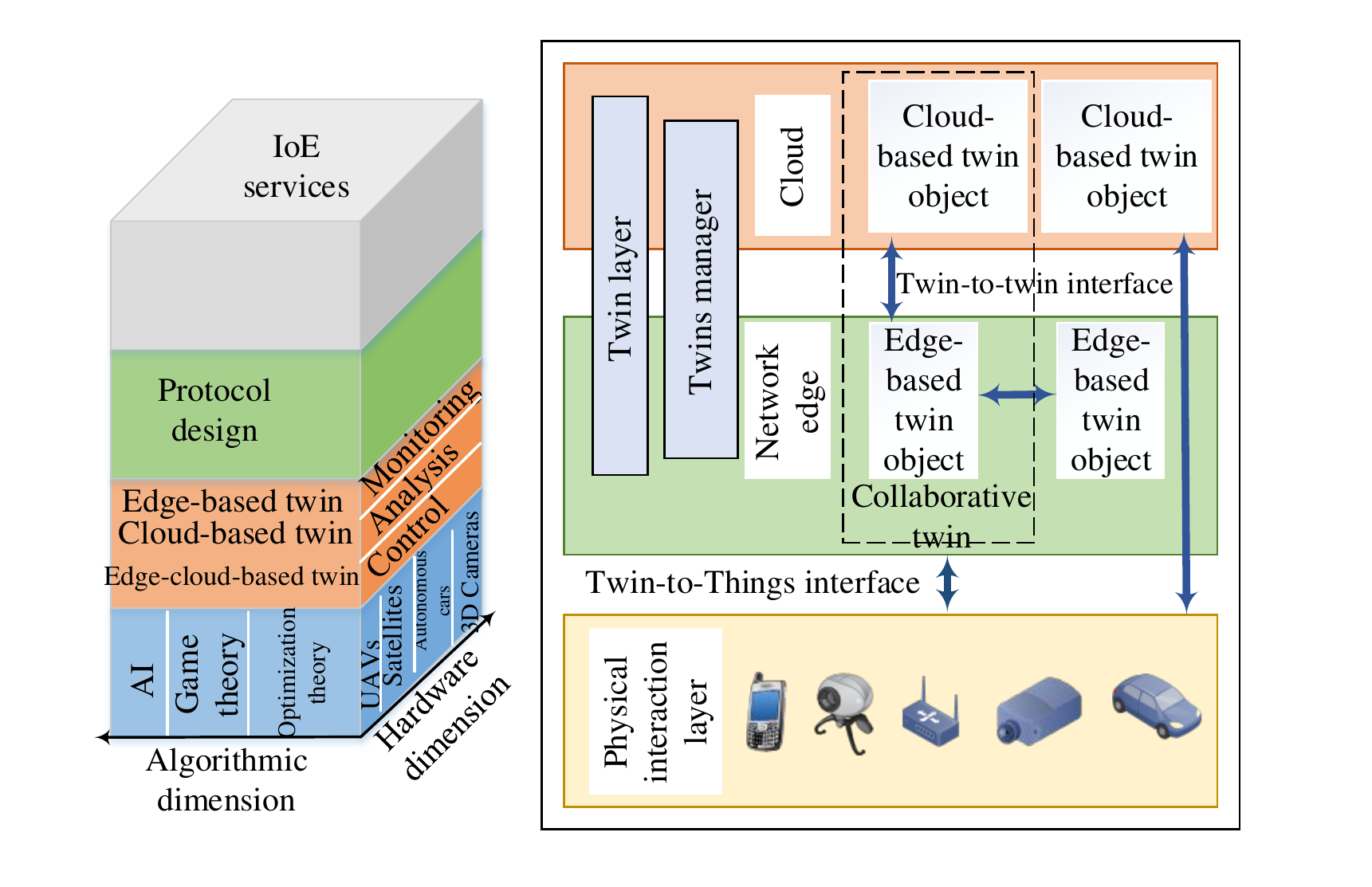}
	\caption{Twin objects deployment trends \cite{khan2021digital}.}
	\label{fig:twindeploymenttrends}
\end{figure}

To benefit from both edge-based twins and cloud-based twins, one can use a hybrid trend that simultaneously deploys twins at network edge and cloud for enabling a certain application. For instance, consider the digital-twin-enabled infotainment system for autonomous cars. One can use caching assisted by a hybrid twin that consists of two twins deployed at both network edge and the cloud. The edge twin will make the caching decisions for infotainment users where latency is stringent. Meanwhile, due to limited storage capacity at the edge, one can use the cloud to cache the information that is less frequent compared to cached information at the edge. To control caching at the cloud, cloud twins can be used. Comparisons of various twins depending on the deployment trends are given in Table~\ref{tab:edgecomparison}. Fig.~\ref{fig:twindeploymenttrends} shows the different ways in which twins can be deployed. Table~\ref{tab:edgecomparison} shows that edge-based twins (i.e., for multiple twin objects) have the highest robustness to failures compared to cloud-based twins and hybrid twins. The reason for this is due to the fact that even if any of the twin objects malfunction due to physical damage or a security attack, the other twin objects will serve the end-user. Similarly, the elasticity of edge-based twins is higher than a cloud-based twin. The reason for this is the remote location of cloud-based twin objects that results in high latency while serving the user. Considering latency as a metric, edge-based twins will have the lowest among the possible deployment trends. The reason for this is due to its nearby location to end-users. However, edge-based twins have low computing power than the cloud-based-twins. Therefore, one must make a tradeoff between the latency and available computing resources for performing a twinning task. Mobility support is also high for edge-based twins due to their nearby nature to end-devices and high context-awareness (i.e., devices locations and edge network information).    \par 

\subsection{Physical End-Devices Design}
To efficiently operate twin-based wireless systems, there is a need to effectively design the end-devices. These end-devices have two main uses, such as actions and training of local models for distributed learning \cite{khan2021digital}. For actions (e.g., optimal transmit power), the devices with the assistance of twins will perform decisions (e.g., optimal transmit power, association, and resource allocation) for optimal performance. For such scenarios, the devices may not require significantly high computing power. However, for the distributed learning case, the devices will require high computing power to learn their local models. In the case of distributed learning-based twins training, the devices will learn local models using their local data. Other than local models training, the devices have to perform additional tasks as well. Training a twin model using a set of devices depends on local learning model architecture, local dataset, and local iterations \cite{khan2020federatededge}. For a fixed local dataset and local model architecture, the computing time is determined by the device operating frequency. The device's local model computing time decreases within an increase in operating frequency. However, this would cause an increase in energy consumption. Note that the energy consumption also depends on the devices design \cite{device_design_1}. Therefore, there is a need to efficiently design the end-devices for twin-based wireless systems. End-devices design can be mainly based on either hardware-based design or software-based design \cite{khan2021federated}. \par      
One can design programmable hardware with high dimensionality for end-devices to enable its general use for various tasks in a digital twin-enabled wireless system \cite{khan2021digital}. Note that high dimensionality causes an increase in computing power because of more generated data. To efficiently enable the hardware design, one can use application-specific manycore processors \cite{ceze2016arch2030}. To implement hardware using manycore processors there is a need to fine-tuning for enabling efficient run-time resource management. Also, the use case has a significant impact on the hardware performance. To enable efficient hardware, Kim \textit{et al.} in \cite{kim2018machine} presented a machine learning-based design for manycore systems. On the other hand, using fixed hardware design, one can propose multiple software designs that can efficiently use the given hardware for performing distributed learning tasks. To do so, we can exploit neural architecture search (NAS) \cite{liu2018progressive,zhou2019epnas,elsken2019neural,pham2018efficient}. In NAS, various possible neural architectures are searched among a search space for certain hardware and the efficient one is selected.  
\subsection{Lessons Learned and Recommendations}
In this section, we devised a taxonomy for twins of wireless aspect. The devised taxonomy considered twin isolation, incentive design, twin object design, twin object prototyping, twin objects deployment trends, physical end-devices design, decoupling, and interfaces design. as parameters. The lessons learned and future recommendations are as follows.
\begin{itemize}
    \item From \cite{khan2021digital,khan2020federatededge,device_design_1,liu2018progressive,zhou2019epnas,elsken2019neural,pham2018efficient}, we learned that there is a need to consider both hardware and software design while designing end-devices. One can jointly consider both hardware designs and neural architectures and selected the one with optimal results. Such a joint design hardware-software co-design will offer more freedom of variations during searching of the optimal hardware and software designs. 
      \item There is a need to efficiently deploy the twin objects at the network edge/cloud. There can be situations, where one service (i.e., deep reinforcement learning-infotainment service) must be served by multiple twin objects (i.e., running agent for caching decision) at edge servers. For such scenarios, one must cost-efficiently deploy the twin objects. To deploy one can use a scheme based on matching theory and optimization theory. 
        
    \item From \cite{Akhan2021survey,tang2020wireless,khan2021digital}, we learned that considerable care must be taken while virtual modeling the physical system/entity. Mathematical models are based on assumptions that may not more accurately follow the real scenarios. On the other hand, one can use experimental modeling that is based on extensive experimentation. However, this approach has the drawback of long design time. Another way can be data driven modeling that is based on training a machine learning model. Again, training a machine learning model may have training time. From the aforementioned facts, there is a need to effectively model the twins while minimizing design time. For instance, one can train a machine learning for twin using efficient architecture that have less training time.  

\end{itemize}
\section{Taxonomy: Wireless for Twins}
Wireless for twins deals with efficient twins signaling for enabling various services. Various parameters of the wireless for twins aspects are air interface design, twin objects access aspects, and security and privacy, as shown in Fig.~\ref{fig:taxonomy}. Air interface will be used for twin signaling. Twin objects access aspects will enable efficient association of twins and physical devices. Finally, there must be effective security and privacy mechanism for twins-based wireless systems.   

\subsection{Air Interface Design}
\label{air interface}
Air interface will be used by digital twin for transfer of learning model updates, data, and control instructions. Meanwhile, there are significant limitations on the availability of wireless resources. Therefore, we must employ efficient communication schemes for wireless transfer. There can be two main aspects for the design of wireless interface for digital twinning, such as frequency band and access scheme. The access schemes can be orthogonal frequency division multiple access (OFDMA), time division multiple access (TDMA), non-orthogonal multiple access (NOMA), etc \cite{6550522}. OFDMA uses orthogonal resource blocks for communication, and thus there will be no interference among them. OFDMA suffers from spectral efficiency and limited users issues. It might not be able to serve the massive number of users. To address this limitation, one can use NOMA. Comparison of NOMA and OFDMA is given in Table~\ref{tab:NOMACOmparison} \cite{wei2016survey}. NOMA uses the whole bandwidth for all users and performs decoding using power levels. NOMA can offer several advantages, such as high spectral density, high connection density, and enhanced user fairness but at the cost of high complexity associated with the receiver. Therefore, a tradeoff must be made between the receiver complexity and spectral efficiency. \\

\begin{table}
 
\caption {Comparison of NOMA and OFDMA \cite{wei2016survey}.} \label{tab:NOMACOmparison} 
  \centering
  \begin{tabular}{p{1cm}p{3.5cm}p{3.5cm}}
    \toprule 
   
     \textbf{Category} &  \textbf{Advantages} & \textbf{Disadvantages}  \\
     \midrule
    
    OFDMA & \begin{itemize} \item Low complexity receiver design \end{itemize} & \begin{itemize} \item Low spectral efficiency \item Limited users  \end{itemize} \\
    
    \midrule
    
    NOMA & \begin{itemize} \item High spectral efficiency \item High connection density \item Enhanced user fairness  \end{itemize}  & \begin{itemize}  \item High complexity of receivers \item Sensitivity to channel uncertainties \end{itemize} \\
    
\bottomrule 
\end{tabular}

\end{table}

On the other hand, one can use many frequency bands for twinning depending on the requirement of the wireless system applications. Lower frequency bands (e.g., medium frequency, high frequency) have been used for various applications (e.g., navigation beacons) that do not have very strict latency requirements. With the emergence of novel applications, such as intelligent transportation systems, human-computer interactions, and XR, among others, the requirement of strict latency becomes one of the most important key performance requirements. To enable low latency communication, there is a need for large bandwidth for communication. To enable such communication, one can use higher frequency bands (i.e., millimeter-wave and terahertz). Although sufficient bandwidth is available at high-frequency bands, there will be sufficiently high attenuation. To address this issue, one can use intelligent reflecting surfaces \cite{zhao2019survey}. Intelligent reflecting surfaces consist of an array of reflecting units with the ability to incur some change independently \cite{basar2019wireless}. Such a change can be either polarization, frequency, amplitude or phase \cite{zhao2019survey}. The main goal of intelligent reflecting surfaces is to enable efficient communication between the transmitter and receiver that does not have a line of sight path. Although intelligent reflecting surfaces have many benefits, it has a few challenges. These challenges are surface design, channel sensing, and estimation, and passive beamforming, among others \cite{gong2020toward}. For channel sensing and estimation in intelligent reflecting surfaces, many works \cite{wang2020compressed, nadeem2019intelligent, liu2020deep} presented several proposals. In \cite{wang2020compressed}, the authors considered estimation for an intelligent reflecting surfaces-enabled millimeter-wave communication system and proposed a compressed sensing-based estimation scheme. Additionally, the authors also extended their work to a multi-antenna system. The work of \cite{wang2020compressed} performed well while minimizing the training overhead. Another work \cite{nadeem2019intelligent} studied modeling and channel estimation for intelligent reflecting surfaces-assisted wireless communication. The works in \cite{wang2020compressed} and \cite{nadeem2019intelligent} may not very effectively perform in a change in SNR scenarios and require retraining. To address this issue, the work \cite{liu2020deep} proposed a hybrid passive/active intelligent reflecting surfaces architecture. They presented a deep denoising neural network-based compressed sensing broadband channel estimation. The focus of the authors was an intelligent reflecting surfaces-assisted millimeter wave system and minimized training overhead. \par

\subsection{Twin Objects access aspects}
There are two main aspects associated with accessing the twin objects, such as (a) twin objects and physical devices association and (b) computing and wireless resource allocation. Twin objects can be deployed at either at cloud or edge server. For edge-based twins, one can apply twins at various network edges that have different access costs (e.g., transmit power, achievable throughput, packet error rate) for physical objects. Therefore, we must properly associate the physical objects with twins. The edge-based twins and physical objects association problem will be a combinatorial problem. Such kinds of problems can be solved using various techniques, such as relaxation-based schemes, heuristic schemes, and matching theory-based solutions \cite{zhou2016energy,zhang2017computing,ahmed2019joint,zhang2017energy,azizi2017joint,bairagi2019matching,kim2019optimal,khan2020self}. The relaxation-based solution first transforms the binary association variable into a continuous variable which can be further solved using various schemes (e.g., convex optimization solver if the relaxed problem is convex). A relaxation-based solution can provide a low complexity solution but at the cost of approximation error (i.e., transforming the binary variables into continuous variables). To avoid the approximation error in relaxation-based schemes, one can use heuristic schemes \cite{khan2020self}. However, heuristic schemes have high computational complexity. Therefore, to address the limitations of both heuristic and relaxation-based schemes, one can use matching theory-based solutions that can offer effectively association between twins and the physical objects with reasonable complexity \cite{bairagi2019matching,kim2019optimal}. \par        
Other than association, there is a need to efficiently allocate wireless resources to physical devices for communication with twin objects. Additionally, computing resources are required for twin processing. There are two main resources in digital twin-enabled wireless systems, such as wireless/wired resources for signaling and computing resources. The computing resources can be used for various tasks. These tasks are running machine learning models at edge/cloud in case of centralized learning or physical devices for distributed learning. Additionally, computing resources are required for blockchain consensus algorithms and analysis of the virtual twin model prior to applying for real-time applications. The computing resource for a machine learning task of size $T_{s}$ units within time $T_{comp}$ for $I$ iterations can be given by.
            \begin{equation}
            \label{eq:3.22}
		    \begin{aligned}
		    f_{n}=I\left(\frac{T_{s}}{T_\textrm{comp}}\right).
		    \end{aligned}
		    \end{equation}\\
More computing resources are required for running the twin model within less time, and vice versa. Additionally, the size of the twin model along with learning model iterations proportionally affect the requirement of computing resources. Therefore, we must make a tradeoff between the task computing time and computing resource requirement. On the other hand, computing resource requirement for blockchain consensus strictly depends on the type of consensus algorithm and network size. \par   
Other than computing resources, communication resources are required for the signaling and training of distributed twin models. Communication resources can be wireless access network resources and core network resources. Generally, the core network delay is significantly small due to high-speed optical backhaul links. Wireless access network resources are limited, therefore, there is a need to efficiently allocate these resources for twinning. The tasks, such as distributed twin models training, twin signaling, and wireless blockchain miners will use communication resources. The required communication resources are dependent on the twinning task. For distributed twin models training, wireless resources are required for sharing the learning model updates between the end-devices and the aggregation server. For twin signaling, wireless resources are required for the transfer of control signals (e.g., SDN controllers signaling). On the other hand, blockchain miners will use wireless resources for sharing of blocks during consensus algorithm. All of the aforementioned use of wireless resources require efficient allocation of wireless resources. Generally, the wireless resource allocation problem is a combinatorial problem that can be solved using various ways, such as heuristic algorithms \cite{khan2020self}, relaxation-based schemes \cite{afolabi2012multicast}, matching theory-based schemes \cite{di2015radio,leanh2017matching,zakeri2019joint}.          

\begin{figure*}[!t]
	\centering
	\captionsetup{justification=centering}
	\includegraphics[width=18cm, height=10cm]{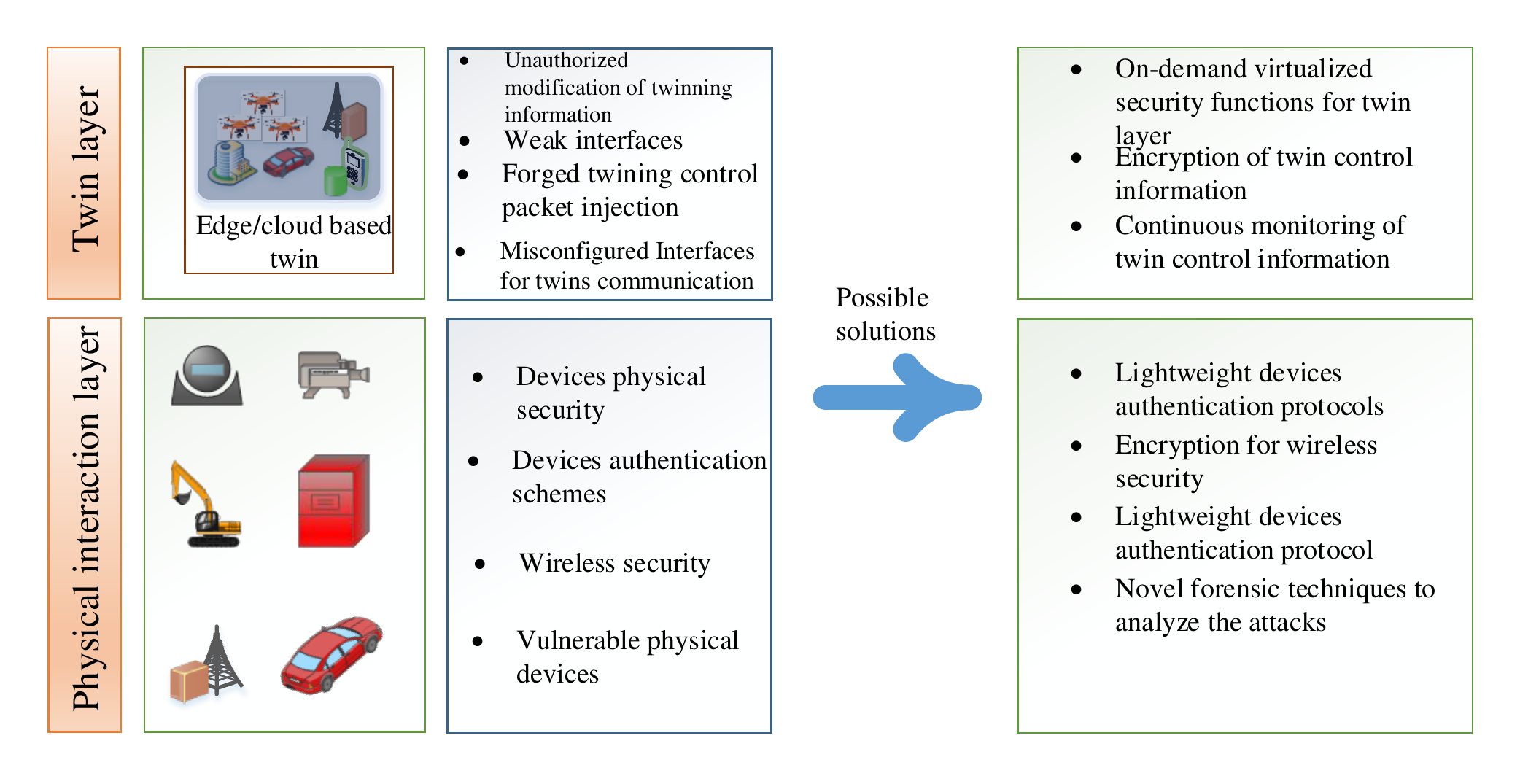}
	\caption{Security attacks in digital twin-based wireless systems.}
	\label{fig:security_attacks}
\end{figure*}

\subsection{Security and Privacy}
\label{security and privacy}
Security in digital twins and wireless systems can be mainly categorized into two types, such as devices physical security and interfaces security. Devices physical security is challenging due to their large number and distributed nature \cite{khan2020edge}. Therefore, one must employ effective authentication schemes to avoid the unauthorized access to devices, edge/cloud servers, and blockchain miners, among others. On the other hand, we must use effective security mechanisms for interfaces in digital twin-based wireless systems. Interfaces in digital twin-based wireless system can be wireless interface (e.g., radio access network), application interface (i.e., smart phone application access screen security), and wired interface (e.g., wired core network). Furthermore, the technologies, such as SDN and NFV used by the digital twin for efficient control of the underlying physical devices arise new security challenges. Overview of various security attacks with their possible solutions for digital twin based wireless system is given in Fig.~\ref{fig:security_attacks}. In SDN control plane, the secruity threats are vulnerable network controllers, forged control packet injection, misconfigured policy enforcement, and weak network devices authentication \cite{chica2020security}. Misconfigured interfaces (e.g., twin to twin interface) and protocols (twin packets routing protocols) results in various security vulnerability \cite{kendall1999database}. Weak or improper authentication schemes and plan text channels may lead to security attacks. Therefore, one must employ effective encryption/decryption schemes to avoid the security attacks. \par

\begin{figure*}[!t]
	\centering
	\captionsetup{justification=centering}
	\includegraphics[width=16cm, height=7cm]{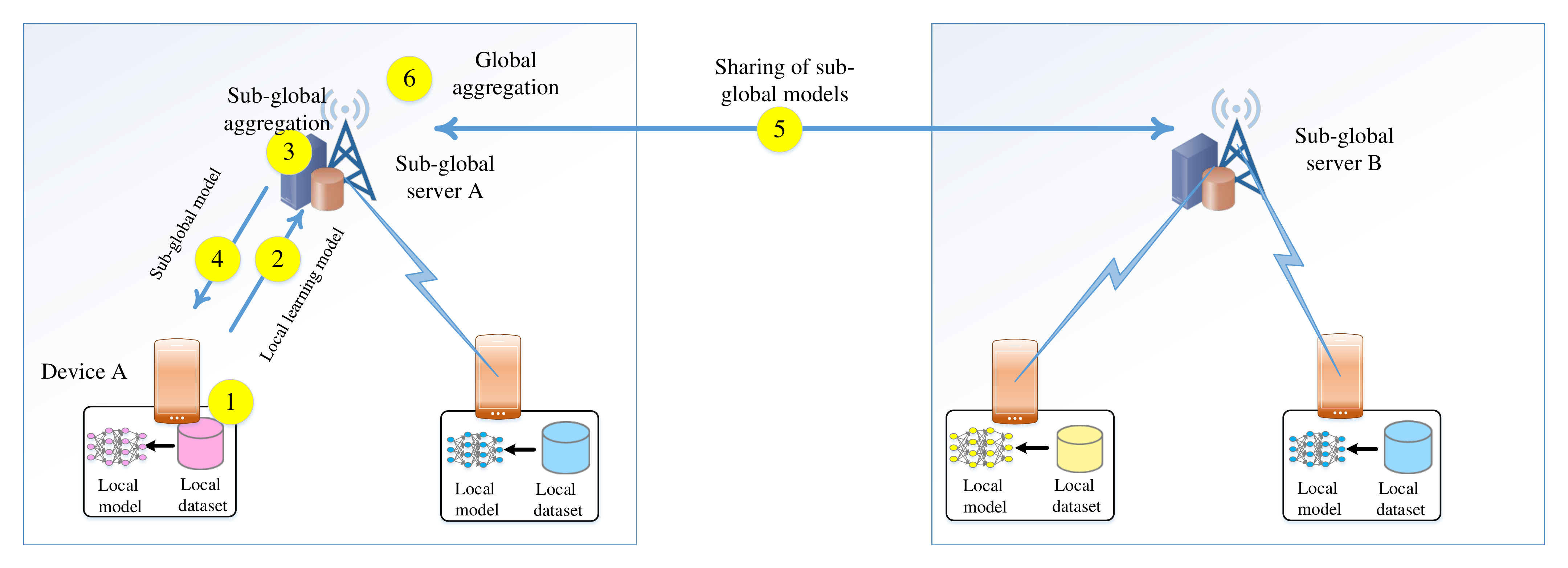}
	\caption{Dispersed federated learning overview.}
	\label{fig:dfl}
\end{figure*}

\begin{figure}[!t]
	\centering
	\captionsetup{justification=centering}
	\includegraphics[width=8cm, height=7cm]{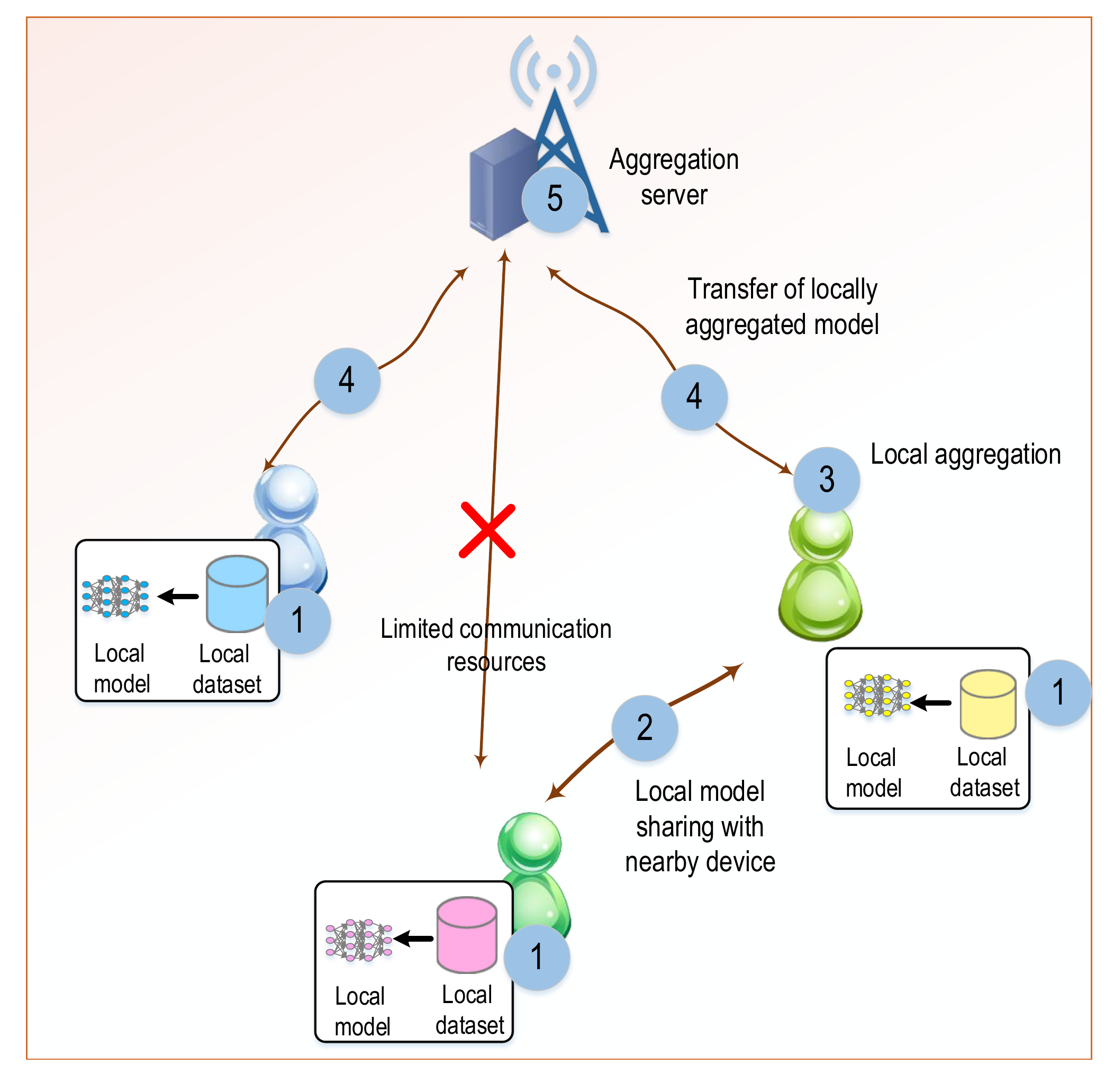}
	\caption{Collaborative federated learning.}
	\label{fig:cfl}
\end{figure}
On the other hand, there may a loss of privacy during training twin models. In the case of centralized training, all the end-devices data is transferred to a centralized server for training and thus results in end-devices privacy leakage. To address this issue, one can train twin models in a distributed manner that is based on on-device training. The local models trained at devices are sent to the edge/cloud for aggregation to yield a global model. The global model is shared with the end-devices again for updating their local models. This process of learning distributed twin models takes place iteratively. Although end-devices do not send their data to the remote cloud/edge server for training, they still require privacy preservation techniques. A malicious aggregation server can infer the end-device sensitive information using their local learning model updates, and thus results in privacy leakage \cite{khan2021federated,khan2021socially,khan2021dispersed}. To address this issue, one can use various schemes, such as differential privacy and homomorphic encryption-based schemes. In differential privacy, a noise is added to the local learning model updates prior to sending them to the aggregation server. Although differential privacy can enhance privacy preservation, it will be at the cost of slowing the global learning convergence \cite{khan2020federatededge}. To avoid this issue, one can use homomorphic encryption that is based on encrypting the local learning models prior to sending them to the aggregation server. Similar to differential privacy, homomorphic encryption works at the cost of communication overhead \cite{khan2021dispersed}. Therefore, a tradeoff must be made the overhead and privacy preservation. Other than differential privacy and homomorphic encryption, few works \cite{yang2020federated, liu2020over,jiang2019over} proposed over-the-air computation. In over-the-air computation, the channel noise is considered a differential privacy noise for preserving the distributed learning privacy. On the other hand, the work in \cite{khan2021dispersed} proposed a dispersed federated learning scheme that is based on computing sub-global models within groups, as shown in Fig.~\ref{fig:dfl}. The computation of the sub-global model is performed iteratively. Then, the sub-global models are shared among each other and finally, the global model is computed. Note that dispersed federated learning offers enhanced privacy protection compared to traditional federated learning. In traditional federated learning, a malicious aggregation server can infer some of the devices' sensitive information from their learning model updates. In contrast to traditional federated learning, a sub-global aggregation server can infer, but a global server can not infer. Therefore, one can say that dispersed federated learning can offer better privacy preservation than traditional federated learning. Another work \cite{chen2020wireless} presented a collaborative federated learning scheme that is based on local aggregation at end-devices, as shown in Fig.~\ref{fig:cfl}. Such local aggregation is performed due to communication resources constraints. The global aggregation can infer the devices' sensitive information from the learning model updates (excluding locally aggregated learning models). However, it is very difficult for the aggregation server to infer the devices' local information from the locally aggregated learning models. Therefore, one can say that collaborative federated learning can offer better privacy preservation compared to traditional federated learning \cite{9236971}.   \par

\subsection{Lessons Learned and Recommendations}
In this section, a taxonomy of twins wireless for twins is devised. We considered three parameters, such as twin objects access aspects, air interface design, security and privacy. The lessons learned and recommendations are as follows.
\begin{itemize}
    \item We learned from \cite{khan2020edge, chica2020security,kendall1999database,khan2021dispersed} that effective security schemes must be applied to digital twin-based wireless systems. Digital twin-based wireless systems will offer many benefits but will be prone to variety of security attacks as discussed in Section~\ref{security and privacy}. Prior to applying to security schemes, one can propose effective forensic schemes to investigate the attacks in digital twin-based system. Next to investigation of security attacks, we can propose efficient security attacks. For devices, one must propose lightweight and effective authentication schemes. For wireless transfer of data and control signals, one can use efficient and effective encryption schemes. Twin-based wireless systems have a layered architecture that communicate with each other using interfaces (e.g., twin object to twin object interface and twin object to physical object interface). There may be configured interfaces for twinning that will result in security attack. Additionally, forged twinning control instructions will significantly degrade the performance of digital twin-based systems. Therefore, there is a need to propose novel security schemes for digital twinning control information transfer.
  
    \item To deploy multiple twin objects for various services, there is a need to efficiently manage computing resource at the network edge/cloud. Different twins (e.g., XR twin, healthcare twin) have different computing resource requirements, therefore, deploying twins with heterogeneous requirements need careful design considerations. We must effectively allocate computing resource among various twins. Additionally, there may be different among twins deployment. For instance, healthcare twins might have more priority compared to infotainment twins. Therefore, computing resource allocation must also take int account the priority of various twins. 

    \item Digital-twin-based system have a wide variety of players (i.e., end-devices, edge/cloud servers, SDN switches). Mostly, the backup power of end-devices has limitations. Therefore, there is a need to propose energy efficient algorithms. For instance, one can propose energy efficient association of physical end-devices with the edge server running twin objects. The energy efficiency for association can be obtained by optimizing the transmit power allocation while performing association.  
  
\end{itemize}


\section{Open Challenges}
This section presents a few open challenges with their possible guidelines. Previous surveys and tutorials have discussed open challenges, such as standardization issues, security and privacy, government regulations for medical twins, accurate representation of digital twins, technical limitations, barriers to blockchain applications in digital twins, and data related issues. Our work lists novel challenges as given in Table~\ref{tab:challengecompariso}. 

\begin{table}[]
\caption {Summary of existing surveys and tutorials open research challenges.} \label{tab:challengecompariso} 
\begin{center}
\begin{tabular}{p{2.2 cm}p{5.7cm}}
\toprule 
    \textbf{Reference}   & \textbf{Challenges} \\ \midrule
Minerva \textit{et al.}, \cite{twin_pro_ieee} & Standardization, scalability, composability, and business model. \\ \midrule

Wu \textit{et al.},~\cite{wu2021digital} & Security vulnerability, privacy leakage, cost-effective solutions, and two-way real-time interaction. 
  \\ \midrule
Barricelli \textit{et al.},~\cite{barricelli2019survey} & Ethical issues, security and privacy, cost of development, equally distributed wealth, government regulations for medical twins, and technical limitations.\\ \midrule
Yaqoob \textit{et al.},~\cite{yaqoob2020blockchain} & Accurately representing an object and affordability of digital twins, ethical, legal, and societal issues, cybersecurity, and barriers to Blockchain Adoption in digital twins.\\ \midrule

Suhail \textit{et al.},~\cite{suhail2021blockchain} & Digital twin representation, data related issues, expenditure on infrastructure.\\ \midrule
Khan \textit{et al.},~\cite{khan2021digital} & Isolation between digital twin-based services, mobility management for edge-based twins, digital twin forensics.\\ \midrule
Our Tutorial & Dynamic twins, interoperability for twins migration, twins prototyping of physical objects, incentive mechanism for twinning, and efficient twin objects chaining.\\

\bottomrule 
\end{tabular}
\end{center}
\end{table}

\begin{table*}
\caption {Summary of the research challenges and their guidelines.} \label{tab:challenges} 
  \centering
  \resizebox{\textwidth}{!}{
  \begin{tabular}{p{3cm}p{3.5cm}p{5cm}p{5cm}}
    \toprule 
   
     \textbf{Challenges} & \textbf{Taxonomy relevancy} &  \textbf{Causes} & \textbf{Guidelines} \\
     \midrule
     \textbf{Dynamic twins} & Twin objects prototyping  &\begin{itemize} \item Physical objects/system dynamic states \item Long design time of new twin objects   \end{itemize} & \begin{itemize} \item  Centralized machine learning-based twins \item Distributed learning-based twins  \end{itemize}\\
     \midrule

     \textbf{Interoperability for twin objects migration}  & Twin objects deployment trends  & \begin{itemize} \item End-devices mobility \item Strict latency constraints of the various services \end{itemize} & \begin{itemize} \item Open cloud/edge computing interface based design \item Similar architecture for edge servers running the twin objects.   \end{itemize} \\
    \midrule
     \textbf{True Prototyping of Physical Objects} & Twin objects prototyping    & \begin{itemize} \item Accurate estimation of twin objects measure \item Dynamic nature of the physical systems    \end{itemize} & \begin{itemize} \item Experimental modeling \item 3D modeling \item Data driven modeling \end{itemize} \\
    \midrule
    \textbf{Incentive Mechanisms for Twinning} & Incentive design & \begin{itemize} \item End-devices consume their resources for training a distributed twin model. \item Miners also perform mining for managing twin object pretrained models. \item Edge/cloud servers require incentives for running twin objects of various services.  \end{itemize} & \begin{itemize} \item  Game theory-based incentive mechanism \item Contract theory-based incentive mechanism \item Auction-based incentive mechanism \end{itemize} \\
    \midrule
    \textbf{Twinning forensics and security}  & Security and privacy  & \begin{itemize} \item Wide variety of players are susceptible to security attacks \item Different players (e.g., edge server, routers) have different architecture  \end{itemize} & \begin{itemize} \item Video-based forensics schemes \item Blockchain-based forensics schemes \item Mobility-aware forensics schemes   \end{itemize} \\
    \midrule
    \textbf{Efficient Twin objects chaining} & Twin objects access aspects & \begin{itemize} \item Design a new twin is computationally expensive \item Long design time for training, testing, and validation for newly designed twins \item High cost associated with new twins design  \end{itemize} & \begin{itemize} \item Mathematical optimization-based schemes \item Game theoretic schemes \item Machine learning-enabled schemes \end{itemize}  \\
\bottomrule 
\end{tabular}
}
\end{table*}

\subsection{Dynamic Twins}
{\em How does one enable twins to be reused for controlling various physical devices?} Re-usability of twin objects is one of the main features that can make digital twins promising for use in wireless systems. Designing twin objects require significant efforts to make them an exact replication of physical object/ phenomenon with energy and computationally efficiency. Therefore, it will be highly desirable to make the twins reusable for future use. There is a need to make the twins general for their use for various services. The general twins can be designed using machine learning. One can train a twin machine learning model for a more general dataset to enable its applicability to different services. However, training a twin machine learning model for general data might not perform well \cite{emmert2020introductory,pouyanfar2018survey}. Additionally, the selection of the machine learning model must be made carefully. For small datasets, generally, machine learning models with low complexity are desirable than the models with high complexity. Therefore, there must effective selection of the twin machine learning models for making it more general. Such a general dynamic twin model can be based on either centralized machine learning or distributed machine learning (e.g., federated learning).

\subsection{Interoperability for Twin Objects Migration}
{\em How does one enable a seamless operation of end-devices served by edge-based twin objects?} Mostly, end-devices in wireless systems are mobile. For instance, one device connected to a small cell base station equipped with an edge server running twin objects may move to the coverage area of another base station. There can be two ways to serve the mobile device. One can be to connect to the existing edge server through the core network via a newly associated base station. However, this approach will suffer from inherent latency that might not be desirable due to strict latency constraints of most of the IoE applications. To address this issue, one can migrate the twin object to the newly associated small cell base station. Twin objects based on virtual machines can be migrated dynamically depending on the mobile device location \cite{masdari2016overview}. One can use machine learning schemes to enable the efficient migration of twin object-based virtual machines. However, it must be noted that transferring a virtual machine from one edge/cloud server to another may face interoperability issues. Two edge/cloud servers must be designed interoperable to ease migration of virtual machines running twin objects to tackle the mobility of end-devices. One can use a common architecture to enable easier migration, such as unified cloud interface/cloud broker, enterprise cloud Orchestration platform/orchestration layer, and open cloud computing interface \cite{thakur2015interoperability}.

\subsection{True Prototyping of Physical Objects}
{\em How do we truly prototype the physical object attributes (e.g., features, data, actions, and events) into twin objects for various applications?} It is necessary to estimate the measurable aspects of the physical objects for twin modeling. However, it is challenging to accurately measure the aspects of a physical system. For instance, it is difficult to measure the aspects of the human body using wearable for healthcare \cite{twin_pro_ieee}. During modeling of a physical object, one can focus on a few parameters more, otherwise, the complexity will be very high. Meanwhile, there are various dynamic phenomena (e.g., wireless channels) in wireless systems that are possible to be exactly determined. Various modeling schemes are experimental modeling, three-dimensional modeling, and data-driven modeling \cite{rasheed2019digital}. Experimental modeling involves full-scale experimentation for understanding a physical phenomenon. Based on the experimentation, one can find the parameters that are difficult to find directly using techniques (e.g., correlations). In three-dimensional modeling, the goal is to develop mathematical models of physical objects using various techniques (e.g., 3D scanning). However, exact representation using a mathematical model is challenging. On the other hand, data-driven modeling uses data for deriving the functional form of physical objects. Machine learning can be used to model physical objects using data. However, proper selection and training of a machine learning model is a challenge and needs careful considerations.

\subsection{Incentive Mechanisms for Twinning}
{\em How does one motivate various players of a twin-based wireless system for successful and effective operation?} In digital twinning, a variety of players, such as edge/cloud servers, miners, end-devices, and network operators, interact with each other to enable wireless services. These players interact to perform various tasks, such as pretraining of twin models, twin operation, and mining for management of twin pre-trained models. For pretraining of twin models using distributed machine learning, there will be two main players, such as end-devices and edge/cloud servers. End-devices compute their local model and expect a monetary incentive for their contributions towards learning of a global model. To design an incentive for such a scenario, one can use the Stackelberg game that follows leader-follower fashion. The edge/cloud server will act as a leader and devices will follow it. The utility of the edge/cloud server can be a function of global accuracy and the end-devices utility can be the difference between cost (i.e., communication and computation) and reward (monetary incentive) \cite{pandey2019crowdsourcing}. Furthermore, one can also use contract theory and auction theory for design incentive for pretraining of twin models \cite{le2020auction,8832210}. On the other hand, one can also design incentives for network operators using game theory, contract theory, and auction theory. The network operator and users can be two players. The network operator will aim to maximize its profit by serving more users while fulfilling their requirements. The end-users will try to improve their performance in terms of throughput.     

\subsection{Twinning Forensics and Security}
{\em How do we investigate and take necessary steps to counter the security attacks in a digital twin-enabled wireless system?} Digital twin-based wireless systems will involve a wide variety of players, such as edge/cloud computing servers, end-devices, various interfaces (e.g., twin to device and twin to twin), and twin objects, among others. The digital twin-based wireless system will have a different architecture than the existing wireless systems. Therefore, they are prone to various kinds of novel attacks (e.g., twin object attacks and twin to twin interface attacks) in addition to existing attacks (e.g., a man in the middle attack). Therefore, one must apply effective forensic schemes to enable the successful operation of twin-based wireless systems. Forensic techniques for a digital twin-based system can be blockchain-based forensics schemes \cite{atlam2018blockchain} and video-based evidence analysis \cite{stoyanova2020survey}. On the other hand, there may be challenges in implementing forensics schemes due to the mobility of nodes. To address this issue, one can propose mobility-aware forensics schemes for digital twins-based wireless systems. As forensics enables to get information about the security attacks, therefore, there is a need to propose effective security mechanisms to meet the security demands of digital twin-based systems. The security scheme can be designed for a complex digital system depending on the point of interest. For instance, SDN and NFV used for decoupling in the digital twin-based wireless system needs different security mechanisms compared to the edge/cloud servers running the twin objects. Similarly, for twin signaling, one can use encryption/decryption schemes.    

\subsection{Efficient Twin Objects Chaining}
{\em How do we chain various twins to enable a service/ wireless system functions for efficient operation?} Generally, designing a twin object requires extensive efforts and time. Therefore, designing novel twin objects for new services will require significant amount of time. For instance, twin-based AR service may requires multiple twin objects deployed at network edge. One way can be to design a novel twin objects for deployment at the network edge. However, this approach will cost in terms of time and efforts. More feasible way can be to reuse the existing twins to enable a complete twin-based AR service. However, combining multiple twins for enabling a service may suffer from high communication and computing cost. Selecting a number of twin objects among a set of available objects is challenging. Every twin object at the edge/cloud is characterized by a certain cost (e.g., latency). For instance, similar twin objects running at edge and cloud server has different latencies. Additionally, the available computing power can also be different depending on the location of twin objects. We must take into account all the factors while chaining twin objects for a certain service/function. Such chaining of twin objects can be based on optimization theory, game theory, heuristic algorithms, and deep reinforcement learning-based schemes. Typically, heuristic algorithms have high computing complexity, and thus may not be suitable for use. One can use optimization and game theory-based schemes. However, there may be a few twin object chaining problems that can not exactly be modeled via mathematical optimization. For such kinds of problems, one can use deep reinforcement learning-based schemes.   

\section{Conclusions and Future Prospects}
We have presented a comprehensive tutorial on digital twins and wireless systems. We presented key design aspects and a high-level framework for a digital-twin-based wireless system. Also, we outlined currently available digital twinning frameworks. Additionally, a comprehensive taxonomy is devised using various parameters. Finally, we presented key open research challenges with causes and possible solutions. We concluded that the integration of digital twins and wireless systems is necessary for enabling IoE applications. Proactive analysis of digital twinning will enable wireless systems to proactively manage the network resources for strict latency applications (e.g., XR). Furthermore, the reusability of generalized digital twins will make them attractive for use in many new emerging applications. \par
As a future prospect, we believe that digital twins will be one of the most promising technology for $6$G and beyond wireless systems. $6$G services will have diverse requirements for novel applications (e.g., human-computer interaction, XR). To meet such kind of diverse requirements, we will redesign the existing wireless systems. These requirements are the user-defined quality of physical experiences, extremely low latency, and ultra-reliability, among others. Enabling a wireless system to meet the aforementioned requirements needs proactive analysis and machine learning-based schemes. The digital twin will enable us to proactively analyze the system and train effective machine learning models. The pre-trained twin machine learning models will enable the wireless system to make on-demand decisions related to the operation of wireless applications in response to demand from users. Furthermore, one can further train the pre-trained twin models to more effectively incorporate the newly generated data. \par

\bibliographystyle{IEEEtran}
\bibliography{Database}

\begin{thebibliography}{100}
\providecommand{\url}[1]{#1}
\csname url@samestyle\endcsname
\providecommand{\newblock}{\relax}
\providecommand{\bibinfo}[2]{#2}
\providecommand{\BIBentrySTDinterwordspacing}{\spaceskip=0pt\relax}
\providecommand{\BIBentryALTinterwordstretchfactor}{4}
\providecommand{\BIBentryALTinterwordspacing}{\spaceskip=\fontdimen2\font plus
\BIBentryALTinterwordstretchfactor\fontdimen3\font minus
  \fontdimen4\font\relax}
\providecommand{\BIBforeignlanguage}[2]{{%
\expandafter\ifx\csname l@#1\endcsname\relax
\typeout{** WARNING: IEEEtran.bst: No hyphenation pattern has been}%
\typeout{** loaded for the language `#1'. Using the pattern for}%
\typeout{** the default language instead.}%
\else
\language=\csname l@#1\endcsname
\fi
#2}}
\providecommand{\BIBdecl}{\relax}
\BIBdecl

\bibitem{al2015internet}
A.~Al-Fuqaha, M.~Guizani, M.~Mohammadi, M.~Aledhari, and M.~Ayyash, ``Internet
  of things: A survey on enabling technologies, protocols, and applications,''
  \emph{IEEE communications surveys \& tutorials}, vol.~17, no.~4, pp.
  2347--2376, Fourthquarter, 2015.

\bibitem{atzori2010internet}
L.~Atzori, A.~Iera, and G.~Morabito, ``The internet of things: A survey,''
  \emph{Computer networks}, vol.~54, no.~15, pp. 2787--2805, October 2010.

\bibitem{li20185g}
S.~Li, L.~Da~Xu, and S.~Zhao, ``5g internet of things: A survey,''
  \emph{Journal of Industrial Information Integration}, vol.~10, pp. 1--9, June
  2018.

\bibitem{ge2018big}
M.~Ge, H.~Bangui, and B.~Buhnova, ``Big data for internet of things: a
  survey,'' \emph{Future generation computer systems}, vol.~87, pp. 601--614,
  October 2018.

\bibitem{khan2021digital}
L.~U. Khan, W.~Saad, D.~Niyato, Z.~Han, and C.~S. Hong, ``Digital-twin-enabled
  6g: Vision, architectural trends, and future directions,'' \emph{To appear,
  IEEE Communication Magazine}, 2022.

\bibitem{tao2018digital}
F.~Tao, H.~Zhang, A.~Liu, and A.~Y. Nee, ``Digital twin in industry:
  State-of-the-art,'' \emph{IEEE Transactions on Industrial Informatics},
  vol.~15, no.~4, pp. 2405--2415, April 2018.

\bibitem{haag2018digital}
S.~Haag and R.~Anderl, ``Digital twin--proof of concept,'' \emph{Manufacturing
  Letters}, vol.~15, pp. 64--66, January 2018.

\bibitem{IoT_tastiscts_1}
Markets and Markets, ``Internet of thigns market,''
  https://www.marketsandmarkets.com/Market-Reports/internet-of-things-market-573.html,
  [Online; accessed Sept. 22, 2021].

\bibitem{IoT_tastiscts_2}
Statistica, ``Spending on internet of things (iot) in selected countries of the
  asia-pacific region in 2019,''
  https://www.statista.com/statistics/1037118/apac-internet-of-things-spending-by-country/,
  [Online; accessed Sept. 22, 2021].

\bibitem{IoT_tastiscts_3}
------, ``Iot spending worldwide 2019, by country,''
  https://www.statista.com/statistics/1118256/iot-spending-worldwide-by-country/,
  [Online; accessed Sept. 22, 2021].

\bibitem{Twin_tastiscts_1}
Markets and Markets, ``Digital twin market,''
  https://www.marketsandmarkets.com/Market-Reports/digital-twin-market-225269522.html?,
  [Online; accessed Sept. 22, 2021].

\bibitem{Twin_tastiscts_2}
B.~Research, ``Global digital twin market,''
  https://www.marketsandmarkets.com/Market-Reports/digital-twin-market-225269522.html?,
  [Online; accessed Sept. 22, 2021].

\bibitem{twin_pro_ieee}
R.~Minerva, G.~M. Lee, and N.~Crespi, ``Digital twin in the iot context: a
  survey on technical features, scenarios, and architectural models,''
  \emph{Proceedings of the IEEE}, vol. 108, no.~10, pp. 1785--1824, October
  2020.

\bibitem{wu2021digital}
Y.~Wu, K.~Zhang, and Y.~Zhang, ``Digital twin networks: a survey,'' \emph{IEEE
  Internet of Things Journal}, Early access, 2021.

\bibitem{barricelli2019survey}
B.~R. Barricelli, E.~Casiraghi, and D.~Fogli, ``A survey on digital twin:
  definitions, characteristics, applications, and design implications,''
  \emph{IEEE access}, vol.~7, pp. 167\,653--167\,671, November 2019.

\bibitem{yaqoob2020blockchain}
I.~Yaqoob, K.~Salah, M.~Uddin, R.~Jayaraman, M.~Omar, and M.~Imran,
  ``Blockchain for digital twins: Recent advances and future research
  challenges,'' \emph{IEEE Network}, vol.~34, no.~5, pp. 290--298,
  September/October 2020.

\bibitem{suhail2021blockchain}
S.~Suhail, R.~Hussain, R.~Jurdak, A.~Oracevic, K.~Salah, and C.~S. Hong,
  ``Blockchain-based digital twins: Research trends, issues, and future
  challenges,'' \emph{arXiv preprint arXiv:2103.11585}, 2021.

\bibitem{Twin_Def_1}
``Digital twins are mission critical,''
  https://www.ge.com/digital/applications/digital-twin, [Online; accessed May.
  16, 2021].

\bibitem{Twin_Def_2}
``What is digital twin?''
  https://www.twi-global.com/technical-knowledge/faqs/what-is-digital-twin,
  [Online; accessed May. 16, 2021].

\bibitem{Twin_soft_1}
``Compare simulation \& cae software,''
  https://www.g2.com/categories/simulation-cae, [Online; accessed May. 16,
  2021].

\bibitem{Twin_soft_2}
``Best digital twin software,'' https://www.g2.com/categories/digital-twin,
  [Online; accessed May. 16, 2021].

\bibitem{Twin_history_1}
``The history and creation of the digital twin concept,''
  https://www.challenge.org/insights/digital-twin-history/, [Online; accessed
  May. 19, 2021].

\bibitem{Twin_type_1}
Siemens, ``Digital twin,''
  https://www.plm.automation.siemens.com/global/en/our-story/glossary/digital-twin/24465,
  [Online; accessed May. 20, 2021].

\bibitem{Twin_type_4}
XMPRO, ``Digital twins: The ultimate guide,''
  https://xmpro.com/digital-twins-the-ultimate-guide//, [Online; accessed May.
  21, 2021].

\bibitem{Twin_type_2}
Vercator, ``How to make a digital twin: the options, types and outputs,''
  https://info.vercator.com/blog/how-to-make-a-digital-twin-the-options-types-and-outputs,
  [Online; accessed May. 21, 2021].

\bibitem{Twin_type_3}
``The digital twin – an introduction,''
  https://www.tributech.io/blog/the-digital-twin, [Online; accessed May. 21,
  2021].

\bibitem{chen2020joint}
M.~Chen, Z.~Yang, W.~Saad, C.~Yin, H.~V. Poor, and S.~Cui, ``A joint learning
  and communications framework for federated learning over wireless networks,''
  \emph{IEEE Transactions on Wireless Communications}, vol.~20, no.~1, pp.
  269--283, January 2020.

\bibitem{khan2020self}
L.~U. Khan, M.~Alsenwi, Z.~Han, and C.~S. Hong, ``Self organizing federated
  learning over wireless networks: A socially aware clustering approach,'' in
  \emph{International Conference on Information Networking (ICOIN)}.\hskip 1em
  plus 0.5em minus 0.4em\relax IEEE, 2020, pp. 453--458.

\bibitem{han2012game}
Z.~Han, D.~Niyato, W.~Saad, T.~Ba{\c{s}}ar, and A.~Hj{\o}rungnes, \emph{Game
  theory in wireless and communication networks: theory, models, and
  applications}.\hskip 1em plus 0.5em minus 0.4em\relax Cambridge university
  press, 2012.

\bibitem{pires2019digital}
F.~Pires, A.~Cachada, J.~Barbosa, A.~P. Moreira, and P.~Leit{\~a}o, ``Digital
  twin in industry 4.0: Technologies, applications and challenges,'' in
  \emph{2019 IEEE 17th International Conference on Industrial Informatics
  (INDIN)}, vol.~1.\hskip 1em plus 0.5em minus 0.4em\relax IEEE, January 2019,
  pp. 721--726.

\bibitem{rasheed2019digital}
A.~Rasheed, O.~San, and T.~Kvamsdal, ``Digital twin: Values, challenges and
  enablers,'' \emph{arXiv preprint arXiv:1910.01719}, 2019.

\bibitem{elayan2021digital}
H.~Elayan, M.~Aloqaily, and M.~Guizani, ``Digital twin for intelligent
  context-aware iot healthcare systems,'' \emph{IEEE Internet of Things
  Journal}, Early access 2021.

\bibitem{ali20206g}
S.~Ali, W.~Saad, N.~Rajatheva, K.~Chang, D.~Steinbach, B.~Sliwa, C.~Wietfeld,
  K.~Mei, H.~Shiri, H.-J. Zepernick \emph{et~al.}, ``6{G} white paper on
  machine learning in wireless communication networks,'' \emph{arXiv preprint
  arXiv:2004.13875}, 2020.

\bibitem{kumar2019machine}
D.~P. Kumar, T.~Amgoth, and C.~S.~R. Annavarapu, ``Machine learning algorithms
  for wireless sensor networks: A survey,'' \emph{Information Fusion}, vol.~49,
  pp. 1--25, September 2019.

\bibitem{jagannath2019machine}
J.~Jagannath, N.~Polosky, A.~Jagannath, F.~Restuccia, and T.~Melodia, ``Machine
  learning for wireless communications in the internet of things: A
  comprehensive survey,'' \emph{Ad Hoc Networks}, vol.~93, p. 101913, October
  2019.

\bibitem{mahdavinejad2018machine}
M.~S. Mahdavinejad, M.~Rezvan, M.~Barekatain, P.~Adibi, P.~Barnaghi, and A.~P.
  Sheth, ``Machine learning for internet of things data analysis: A survey,''
  \emph{Digital Communications and Networks}, vol.~4, no.~3, pp. 161--175,
  August 2018.

\bibitem{sudharsan2021machine}
B.~Sudharsan and P.~Patel, ``Machine learning meets internet of things: From
  theory to practice,'' 2021.

\bibitem{khan2020edge}
L.~U. Khan, I.~Yaqoob, N.~H. Tran, S.~A. Kazmi, T.~N. Dang, and C.~S. Hong,
  ``Edge-computing-enabled smart cities: A comprehensive survey,'' \emph{IEEE
  Internet of Things Journal}, vol.~7, no.~10, pp. 10\,200--10\,232, October
  2020.

\bibitem{fernando2013mobile}
N.~Fernando, S.~W. Loke, and W.~Rahayu, ``Mobile cloud computing: A survey,''
  \emph{Future generation computer systems}, vol.~29, no.~1, pp. 84--106,
  January 2013.

\bibitem{coutinho2015elasticity}
E.~F. Coutinho, F.~R. de~Carvalho~Sousa, P.~A.~L. Rego, D.~G. Gomes, and J.~N.
  de~Souza, ``Elasticity in cloud computing: a survey,'' \emph{annals of
  telecommunications-annales des t{\'e}l{\'e}communications}, vol.~70, no.~7,
  pp. 289--309, 2015.

\bibitem{khan2019edge}
W.~Z. Khan, E.~Ahmed, S.~Hakak, I.~Yaqoob, and A.~Ahmed, ``Edge computing: A
  survey,'' \emph{Future Generation Computer Systems}, vol.~97, pp. 219--235,
  August 2019.

\bibitem{abbas2017mobile}
N.~Abbas, Y.~Zhang, A.~Taherkordi, and T.~Skeie, ``Mobile edge computing: A
  survey,'' \emph{IEEE Internet of Things Journal}, vol.~5, no.~1, pp.
  450--465, September 2017.

\bibitem{twin_service_reliability}
https://www.altoros.com/blog/digital-twins-for-aerospace-better-fleet-reliability-and-performance/,
  [Online; accessed Sept. 27, 2021].

\bibitem{lun2008coding}
D.~S. Lun, M.~M{\'e}dard, R.~Koetter, and M.~Effros, ``On coding for reliable
  communication over packet networks,'' \emph{Physical Communication}, vol.~1,
  no.~1, pp. 3--20, 2008.

\bibitem{guo2012practical}
Z.~Guo, J.~Huang, B.~Wang, S.~Zhou, J.-H. Cui, and P.~Willett, ``A practical
  joint network-channel coding scheme for reliable communication in wireless
  networks,'' \emph{Ieee transactions on wireless communications}, vol.~11,
  no.~6, pp. 2084--2094, 2012.

\bibitem{hausl2009joint}
C.~Hausl, ``Joint network-channel coding for the multiple-access relay channel
  based on turbo codes,'' \emph{European Transactions on Telecommunications},
  vol.~20, no.~2, pp. 175--181, 2009.

\bibitem{twin_frameowrk2}
``Eclipse ditto,'' https://www.eclipse.org/ditto/, [Online; accessed Sept. 22,
  2021].

\bibitem{MCXtwin_frameowrk}
https://github.com/COMEA-TUAS/mcx-public, [Online; accessed Sept. 24, 2021].

\bibitem{MCXtwin_frameowrk2}
https://research.utu.fi/converis/portal/detail/Publication/66484819?, [Online;
  accessed Sept. 22, 2021].

\bibitem{Mago3D_frameowrk1}
http://mago3d.com/en/index.html, [Online; accessed Sept. 27, 2021].

\bibitem{Mago3D_frameowrk2}
https://callforpapers.2021.foss4g.org/foss4g2021/talk/MJPNQC/, [Online;
  accessed Sept. 27, 2021].

\bibitem{ludwig2020efficient}
K.~Ludwig, A.~Fendt, and B.~Bauer, ``An efficient online heuristic for mobile
  network slice embedding,'' in \emph{2020 23rd Conference on Innovation in
  Clouds, Internet and Networks and Workshops (ICIN)}.\hskip 1em plus 0.5em
  minus 0.4em\relax IEEE, 2020, pp. 139--143.

\bibitem{de2019deep}
S.~De~Bast, R.~Torrea-Duran, A.~Chiumento, S.~Pollin, and H.~Gacanin, ``Deep
  reinforcement learning for dynamic network slicing in ieee 802.11 networks,''
  in \emph{IEEE INFOCOM 2019-IEEE Conference on Computer Communications
  Workshops (INFOCOM WKSHPS)}.\hskip 1em plus 0.5em minus 0.4em\relax IEEE,
  2019, pp. 264--269.

\bibitem{kazmi2020distributedss}
S.~A. Kazmi, A.~Ndikumana, A.~Manzoor, W.~Saad, C.~S. Hong \emph{et~al.},
  ``Distributed radio slice allocation in wireless network virtualization:
  Matching theory meets auctions,'' \emph{IEEE Access}, vol.~8, pp.
  73\,494--73\,507, 2020.

\bibitem{ho2018network}
T.~M. Ho, N.~H. Tran, L.~B. Le, Z.~Han, S.~A. Kazmi, and C.~S. Hong, ``Network
  virtualization with energy efficiency optimization for wireless heterogeneous
  networks,'' \emph{IEEE Transactions on Mobile Computing}, vol.~18, no.~10,
  pp. 2386--2400, 2018.

\bibitem{kazmi2017matching}
S.~M.~A. Kazmi and C.~S. Hong, ``A matching game approach for resource
  allocation in wireless network virtualization,'' in \emph{Proceedings of the
  11th International Conference on ubiquitous information management and
  communication}, 2017, pp. 1--6.

\bibitem{kazmi2019network}
S.~A. Kazmi, L.~U. Khan, N.~H. Tran, and C.~S. Hong, \emph{Network slicing for
  5G and beyond networks}.\hskip 1em plus 0.5em minus 0.4em\relax Springer,
  2019, vol.~1.

\bibitem{khan2020network}
L.~U. Khan, I.~Yaqoob, N.~H. Tran, Z.~Han, and C.~S. Hong, ``Network slicing:
  Recent advances, taxonomy, requirements, and open research challenges,''
  \emph{IEEE Access}, vol.~8, pp. 36\,009--36\,028, February 2020.

\bibitem{chang2018radio}
C.-Y. Chang, N.~Nikaein, and T.~Spyropoulos, ``Radio access network resource
  slicing for flexible service execution,'' in \emph{IEEE INFOCOM 2018-IEEE
  Conference on Computer Communications Workshops (INFOCOM WKSHPS)}.\hskip 1em
  plus 0.5em minus 0.4em\relax IEEE, 2018, pp. 668--673.

\bibitem{datsika2016matching}
E.~Datsika, A.~Antonopoulos, N.~Zorba, and C.~Verikoukis, ``Matching game based
  virtualization in shared lte-a networks,'' in \emph{2016 IEEE Global
  Communications Conference (GLOBECOM)}.\hskip 1em plus 0.5em minus 0.4em\relax
  IEEE, 2016, pp. 1--6.

\bibitem{kazmi2017hierarchical}
S.~A. Kazmi, N.~H. Tran, T.~M. Ho, and C.~S. Hong, ``Hierarchical matching game
  for service selection and resource purchasing in wireless network
  virtualization,'' \emph{IEEE Communications Letters}, vol.~22, no.~1, pp.
  121--124, 2017.

\bibitem{Twin_iso1}
``How a vm is realy isolated?''
  https://vinfrastructure.it/2017/04/vm-really-isolated/, [Online; accessed
  Oct. 19, 2021].

\bibitem{Twin_iso2}
https://docs.vmware.com/en/VMware-Cloud-Director/index.html, [Online; accessed
  Oct. 19, 2021].

\bibitem{bera2017software}
S.~Bera, S.~Misra, and A.~V. Vasilakos, ``Software-defined networking for
  internet of things: A survey,'' \emph{IEEE Internet of Things Journal},
  vol.~4, no.~6, pp. 1994--2008, 2017.

\bibitem{mohammed2020review}
A.~H. Mohammed, R.~M. KHALEEFAH, I.~A. Abdulateef \emph{et~al.}, ``A review
  software defined networking for internet of things,'' in \emph{2020
  International Congress on Human-Computer Interaction, Optimization and
  Robotic Applications (HORA)}.\hskip 1em plus 0.5em minus 0.4em\relax IEEE,
  2020, pp. 1--8.

\bibitem{kalkan2017securing}
K.~Kalkan and S.~Zeadally, ``Securing internet of things with software defined
  networking,'' \emph{IEEE Communications Magazine}, vol.~56, no.~9, pp.
  186--192, 2017.

\bibitem{han2015network}
B.~Han, V.~Gopalakrishnan, L.~Ji, and S.~Lee, ``Network function
  virtualization: Challenges and opportunities for innovations,'' \emph{IEEE
  Communications Magazine}, vol.~53, no.~2, pp. 90--97, 2015.

\bibitem{mijumbi2015network}
R.~Mijumbi, J.~Serrat, J.-L. Gorricho, N.~Bouten, F.~De~Turck, and R.~Boutaba,
  ``Network function virtualization: State-of-the-art and research
  challenges,'' \emph{IEEE Communications surveys \& tutorials}, vol.~18,
  no.~1, pp. 236--262, 2015.

\bibitem{joshi2016network}
K.~Joshi and T.~Benson, ``Network function virtualization,'' \emph{IEEE
  Internet Computing}, vol.~20, no.~6, pp. 7--9, 2016.

\bibitem{lal2017nfv}
S.~Lal, T.~Taleb, and A.~Dutta, ``Nfv: Security threats and best practices,''
  \emph{IEEE Communications Magazine}, vol.~55, no.~8, pp. 211--217, August
  2017.

\bibitem{wang2018survey}
S.~Wang, J.~Xu, N.~Zhang, and Y.~Liu, ``A survey on service migration in mobile
  edge computing,'' \emph{IEEE Access}, vol.~6, pp. 23\,511--23\,528, April
  2018.

\bibitem{hassan2019edge}
N.~Hassan, K.-L.~A. Yau, and C.~Wu, ``Edge computing in 5g: A review,''
  \emph{IEEE Access}, vol.~7, pp. 127\,276--127\,289, August 2019.

\bibitem{hu2015mobile}
Y.~C. Hu, M.~Patel, D.~Sabella, N.~Sprecher, and V.~Young, ``Mobile edge
  computing—a key technology towards 5{G},'' \emph{ETSI white paper},
  vol.~11, no.~11, pp. 1--16, 2015.

\bibitem{karakus2017survey}
M.~Karakus and A.~Durresi, ``A survey: Control plane scalability issues and
  approaches in software-defined networking (sdn),'' \emph{Computer Networks},
  vol. 112, pp. 279--293, January 2017.

\bibitem{amiri2019efficient}
E.~Amiri, E.~Alizadeh, and K.~Raeisi, ``An efficient hierarchical distributed
  sdn controller model,'' in \emph{2019 5th Conference on Knowledge Based
  Engineering and Innovation (KBEI)}.\hskip 1em plus 0.5em minus 0.4em\relax
  IEEE, 2019, pp. 553--557.

\bibitem{sarmiento2021decentralized}
D.~E. Sarmiento, A.~Lebre, L.~Nussbaum, and A.~Chari, ``Decentralized sdn
  control plane for a distributed cloud-edge infrastructure: A survey,''
  \emph{IEEE Communications Surveys \& Tutorials}, Firstquarter 2021.

\bibitem{bannour2017distributed}
F.~Bannour, S.~Souihi, and A.~Mellouk, ``Distributed sdn control: Survey,
  taxonomy, and challenges,'' \emph{IEEE Communications Surveys \& Tutorials},
  vol.~20, no.~1, pp. 333--354, Firstquarter 2017.

\bibitem{togou2019hierarchical}
M.~A. Togou, D.~A. Chekired, L.~Khoukhi, and G.-M. Muntean, ``A hierarchical
  distributed control plane for path computation scalability in large scale
  software-defined networks,'' \emph{IEEE Transactions on Network and Service
  Management}, vol.~16, no.~3, pp. 1019--1031, September 2019.

\bibitem{ahmad2021scalability}
S.~Ahmad and A.~H. Mir, ``Scalability, consistency, reliability and security in
  sdn controllers: A survey of diverse sdn controllers,'' \emph{Journal of
  Network and Systems Management}, vol.~29, no.~1, pp. 1--59, November 2021.

\bibitem{huang2016hybridflow}
S.~Huang, J.~Zhao, and X.~Wang, ``Hybridflow: A lightweight control plane for
  hybrid sdn in enterprise networks,'' in \emph{2016 IEEE/ACM 24th
  International Symposium on Quality of Service (IWQoS)}.\hskip 1em plus 0.5em
  minus 0.4em\relax IEEE, June 2016, pp. 1--2.

\bibitem{fu2015hybrid}
Y.~Fu, J.~Bi, Z.~Chen, K.~Gao, B.~Zhang, G.~Chen, and J.~Wu, ``A hybrid
  hierarchical control plane for flow-based large-scale software-defined
  networks,'' \emph{IEEE Transactions on Network and Service Management},
  vol.~12, no.~2, pp. 117--131, June 2015.

\bibitem{interface_2}
https://trackinno.com/iot/how-iot-works-part-4-user-interface/, [Online;
  accessed Oct. 19, 2021].

\bibitem{interface_1}
https://www.business-standard.com/, [Online; accessed Oct. 19, 2021].

\bibitem{virtual_machine_1}
http://course.ece.cmu.edu/~ece600/fall17/lectures/lecture19.pdf, [Online;
  accessed Oct. 21, 2021].

\bibitem{vM_1}
NetApp, ``Containers vs. virtual machines (vms): What's the difference?''
  https://www.netapp.com/blog/containers-vs-vms/, [Online; accessed Oct. 19,
  2021].

\bibitem{vM_2}
``To containerize or not to containerize, that is the question, or containers
  vs vms: the eternal debate,''
  https://www.mirantis.com/blog/containers-vs-vms-eternal-debate/, [Online;
  accessed Oct. 19, 2021].

\bibitem{Akhan2021survey}
A.~S. Khan and F.~U. Khan, ``A survey of wearable energy harvesting systems,''
  \emph{International Journal of Energy Research}, Early access, 2021.

\bibitem{tang2020wireless}
W.~Tang, M.~Z. Chen, X.~Chen, J.~Y. Dai, Y.~Han, M.~Di~Renzo, Y.~Zeng, S.~Jin,
  Q.~Cheng, and T.~J. Cui, ``Wireless communications with reconfigurable
  intelligent surface: Path loss modeling and experimental measurement,''
  \emph{IEEE Transactions on Wireless Communications}, vol.~20, no.~1, pp.
  421--439, January 2020.

\bibitem{khan2021federated}
L.~U. Khan, W.~Saad, Z.~Han, E.~Hossain, and C.~S. Hong, ``Federated learning
  for internet of things: Recent advances, taxonomy, and open challenges,''
  \emph{IEEE Communications Surveys \& Tutorials}, vol.~23, no.~3, pp.
  1759--1799, June 2021.

\bibitem{pandey2019crowdsourcing}
S.~R. {Pandey}, N.~H. {Tran}, M.~{Bennis}, Y.~K. {Tun}, A.~{Manzoor}, and C.~S.
  {Hong}, ``A crowdsourcing framework for on-device federated learning,''
  \emph{IEEE Transactions on Wireless Communications}, vol.~19, no.~5, pp.
  3241--3256, May 2020.

\bibitem{le2020auction}
T.~H.~T. Le, N.~H. Tran, Y.~K. Tun, Z.~Han, and C.~S. Hong, ``Auction based
  incentive design for efficient federated learning in cellular wireless
  networks,'' in \emph{2020 IEEE Wireless Communications and Networking
  Conference (WCNC)}, 2020, pp. 1--6.

\bibitem{8832210}
J.~{Kang}, Z.~{Xiong}, D.~{Niyato}, S.~{Xie}, and J.~{Zhang}, ``Incentive
  mechanism for reliable federated learning: A joint optimization approach to
  combining reputation and contract theory,'' \emph{IEEE Internet of Things
  Journal}, vol.~6, no.~6, pp. 10\,700--10\,714, September 2019.

\bibitem{khan2021dispersed}
L.~U. Khan, Y.~K. Tun, M.~Alsenwi, M.~Imran, Z.~Han, and C.~S. Hong, ``A
  dispersed federated learning framework for 6g-enabled autonomous driving
  cars,'' \emph{arXiv preprint arXiv:2105.09641}, 2021.

\bibitem{ommunication}
Y.~Liu, Z.~Fang, M.~H. Cheung, W.~Cai, and J.~Huang, ``An incentive mechanism
  for sustainable blockchain storage,'' \emph{arXiv preprint
  arXiv:2103.05866v3}, 2021.

\bibitem{khan2020federatededge}
L.~U. Khan, S.~R. Pandey, N.~H. Tran, W.~Saad, Z.~Han, M.~N. Nguyen, and C.~S.
  Hong, ``Federated learning for edge networks: Resource optimization and
  incentive mechanism,'' \emph{IEEE Communications Magazine}, vol.~58, no.~10,
  pp. 88--93, October 2020.

\bibitem{device_design_1}
``Designing for extreme low power,''
  https://semiengineering.com/designing-for-extreme-low-power/, [Online;
  accessed Nov. 10, 2021].

\bibitem{ceze2016arch2030}
L.~Ceze, M.~D. Hill, and T.~F. Wenisch, ``Arch2030: A vision of computer
  architecture research over the next 15 years,'' \emph{arXiv preprint
  arXiv:1612.03182}, 2016.

\bibitem{kim2018machine}
R.~G. Kim, J.~R. Doppa, and P.~P. Pande, ``Machine learning for design space
  exploration and optimization of manycore systems,'' in \emph{IEEE/ACM
  International Conference on Computer-Aided Design (ICCAD)}.\hskip 1em plus
  0.5em minus 0.4em\relax IEEE, 2018, pp. 1--6.

\bibitem{liu2018progressive}
C.~Liu, B.~Zoph, M.~Neumann, J.~Shlens, W.~Hua, L.-J. Li, L.~Fei-Fei,
  A.~Yuille, J.~Huang, and K.~Murphy, ``Progressive neural architecture
  search,'' in \emph{Proceedings of the European conference on computer vision
  (ECCV)}, 2018, pp. 19--34.

\bibitem{zhou2019epnas}
Y.~Zhou, P.~Wang, S.~Arik, H.~Yu, S.~Zawad, F.~Yan, and G.~Diamos, ``Epnas:
  Efficient progressive neural architecture search,'' \emph{arXiv preprint
  arXiv:1907.04648}, 2019.

\bibitem{elsken2019neural}
T.~Elsken, J.~H. Metzen, and F.~Hutter, ``Neural architecture search: A
  survey,'' \emph{The Journal of Machine Learning Research}, vol.~20, no.~1,
  pp. 1997--2017, March 2019.

\bibitem{pham2018efficient}
H.~Pham, M.~Guan, B.~Zoph, Q.~Le, and J.~Dean, ``Efficient neural architecture
  search via parameters sharing,'' in \emph{International Conference on Machine
  Learning}.\hskip 1em plus 0.5em minus 0.4em\relax PMLR, 2018, pp. 4095--4104.

\bibitem{6550522}
L.~U. Khan, Z.~Sabir, S.~A. Mahmud, and G.~M. Khan, ``Comparison of three
  interpolation techniques in comb-type pilot-assisted channel coded ofdm
  system,'' in \emph{27th International Conference on Advanced Information
  Networking and Applications Workshops}, 2013, pp. 977--981.

\bibitem{wei2016survey}
Z.~Wei, J.~Yuan, D.~W.~K. Ng, M.~Elkashlan, and Z.~Ding, ``A survey of downlink
  non-orthogonal multiple access for 5g wireless communication networks,''
  \emph{arXiv preprint arXiv:1609.01856}, 2016.

\bibitem{zhao2019survey}
J.~Zhao, ``A survey of intelligent reflecting surfaces (irss): Towards 6g
  wireless communication networks with massive mimo 2.0,'' \emph{arXiv preprint
  arXiv:1907.04789}, 2019.

\bibitem{basar2019wireless}
E.~Basar, M.~Di~Renzo, J.~De~Rosny, M.~Debbah, M.-S. Alouini, and R.~Zhang,
  ``Wireless communications through reconfigurable intelligent surfaces,''
  \emph{IEEE access}, vol.~7, pp. 116\,753--116\,773, August 2019.

\bibitem{gong2020toward}
S.~Gong, X.~Lu, D.~T. Hoang, D.~Niyato, L.~Shu, D.~I. Kim, and Y.-C. Liang,
  ``Toward smart wireless communications via intelligent reflecting surfaces: A
  contemporary survey,'' \emph{IEEE Communications Surveys \& Tutorials},
  vol.~22, no.~4, pp. 2283--2314, Fourth Quarter, 2020.

\bibitem{wang2020compressed}
P.~Wang, J.~Fang, H.~Duan, and H.~Li, ``Compressed channel estimation for
  intelligent reflecting surface-assisted millimeter wave systems,'' \emph{IEEE
  Signal Processing Letters}, vol.~27, pp. 905--909, May 2020.

\bibitem{nadeem2019intelligent}
Q.-U.-A. Nadeem, A.~Kammoun, A.~Chaaban, M.~Debbah, and M.-S. Alouini,
  ``Intelligent reflecting surface assisted wireless communication: Modeling
  and channel estimation,'' \emph{arXiv preprint arXiv:1906.02360}, 2019.

\bibitem{liu2020deep}
S.~Liu, Z.~Gao, J.~Zhang, M.~Di~Renzo, and M.-S. Alouini, ``Deep denoising
  neural network assisted compressive channel estimation for mmwave intelligent
  reflecting surfaces,'' \emph{IEEE Transactions on Vehicular Technology},
  vol.~69, no.~8, pp. 9223--9228, August 2020.

\bibitem{zhou2016energy}
Z.~Zhou, K.~Ota, M.~Dong, and C.~Xu, ``Energy-efficient matching for resource
  allocation in d2d enabled cellular networks,'' \emph{IEEE Transactions on
  Vehicular Technology}, vol.~66, no.~6, pp. 5256--5268, June 2016.

\bibitem{zhang2017computing}
H.~Zhang, Y.~Xiao, S.~Bu, D.~Niyato, F.~R. Yu, and Z.~Han, ``Computing resource
  allocation in three-tier iot fog networks: A joint optimization approach
  combining stackelberg game and matching,'' \emph{IEEE Internet of Things
  Journal}, vol.~4, no.~5, pp. 1204--1215, October 2017.

\bibitem{ahmed2019joint}
A.~Ahmed, M.~Awais, T.~Akram, S.~Kulac, M.~Alhussein, and K.~Aurangzeb, ``Joint
  placement and device association of uav base stations in iot networks,''
  \emph{Sensors}, vol.~19, no.~9, p. 2157, 2019.

\bibitem{zhang2017energy}
H.~Zhang, S.~Huang, C.~Jiang, K.~Long, V.~C. Leung, and H.~V. Poor, ``Energy
  efficient user association and power allocation in millimeter-wave-based
  ultra dense networks with energy harvesting base stations,'' \emph{IEEE
  Journal on Selected Areas in Communications}, vol.~35, no.~9, pp. 1936--1947,
  September 2017.

\bibitem{azizi2017joint}
A.~Azizi, N.~Mokari, and M.~R. Javan, ``Joint radio resource allocation, 3d
  placement and user association of aerial base stations in iot networks,''
  \emph{arXiv preprint arXiv:1710.05315}, 2017.

\bibitem{bairagi2019matching}
A.~K. Bairagi, M.~S. Munir, M.~Alsenwi, N.~H. Tran, and C.~S. Hong, ``A
  matching based coexistence mechanism between embb and urllc in 5g wireless
  networks,'' in \emph{Proceedings of the 34th ACM/SIGAPP Symposium on Applied
  Computing}, 2019, pp. 2377--2384.

\bibitem{kim2019optimal}
K.~Kim and C.~S. Hong, ``Optimal task-uav-edge matching for computation
  offloading in uav assisted mobile edge computing,'' in \emph{20th
  Asia-Pacific Network Operations and Management Symposium (APNOMS)}.\hskip 1em
  plus 0.5em minus 0.4em\relax IEEE, 2019, pp. 1--4.

\bibitem{afolabi2012multicast}
R.~O. Afolabi, A.~Dadlani, and K.~Kim, ``Multicast scheduling and resource
  allocation algorithms for ofdma-based systems: A survey,'' \emph{IEEE
  Communications Surveys \& Tutorials}, vol.~15, no.~1, pp. 240--254, First
  Quarter 2012.

\bibitem{di2015radio}
B.~Di, S.~Bayat, L.~Song, and Y.~Li, ``Radio resource allocation for downlink
  non-orthogonal multiple access (noma) networks using matching theory,'' in
  \emph{IEEE global communications conference (GLOBECOM)}.\hskip 1em plus 0.5em
  minus 0.4em\relax IEEE, 2015, pp. 1--6.

\bibitem{leanh2017matching}
T.~LeAnh, N.~H. Tran, W.~Saad, L.~B. Le, D.~Niyato, T.~M. Ho, and C.~S. Hong,
  ``Matching theory for distributed user association and resource allocation in
  cognitive femtocell networks,'' \emph{IEEE Transactions on Vehicular
  Technology}, vol.~66, no.~9, pp. 8413--8428, September 2017.

\bibitem{zakeri2019joint}
A.~Zakeri, M.~Moltafet, and N.~Mokari, ``Joint radio resource allocation and
  sic ordering in noma-based networks using submodularity and matching
  theory,'' \emph{IEEE Transactions on Vehicular Technology}, vol.~68, no.~10,
  pp. 9761--9773, October 2019.

\bibitem{chica2020security}
J.~C.~C. Chica, J.~C. Imbachi, and J.~F.~B. Vega, ``Security in sdn: A
  comprehensive survey,'' \emph{Journal of Network and Computer Applications},
  vol. 159, p. 102595, June 2020.

\bibitem{kendall1999database}
K.~K.~R. Kendall, ``A database of computer attacks for the evaluation of
  intrusion detection systems,'' Ph.D. dissertation, Massachusetts Institute of
  Technology, 1999.

\bibitem{khan2021socially}
L.~U. Khan, Z.~Han, D.~Niyato, and C.~S. Hong,
  ``Socially-aware-clustering-enabled federated learning for edge networks,''
  \emph{IEEE Transactions on Network and Service Management}, vol.~18, no.~3,
  pp. 2641--2658, September 2021.

\bibitem{yang2020federated}
K.~Yang, T.~Jiang, Y.~Shi, and Z.~Ding, ``Federated learning via over-the-air
  computation,'' \emph{IEEE Transactions on Wireless Communications}, vol.~19,
  no.~3, pp. 2022--2035, March 2020.

\bibitem{liu2020over}
W.~Liu, X.~Zang, Y.~Li, and B.~Vucetic, ``Over-the-air computation systems:
  Optimization, analysis and scaling laws,'' \emph{IEEE Transactions on
  Wireless Communications}, vol.~19, no.~8, pp. 5488--5502, August 2020.

\bibitem{jiang2019over}
T.~Jiang and Y.~Shi, ``Over-the-air computation via intelligent reflecting
  surfaces,'' in \emph{2019 IEEE Global Communications Conference
  (GLOBECOM)}.\hskip 1em plus 0.5em minus 0.4em\relax IEEE, 2019, pp. 1--6.

\bibitem{chen2020wireless}
M.~Chen, H.~V. Poor, W.~Saad, and S.~Cui, ``Wireless communications for
  collaborative federated learning,'' \emph{IEEE Communications Magazine},
  vol.~58, no.~12, pp. 48--54, January 2020.

\bibitem{9236971}
U.~Majeed, L.~U. Khan, and C.~S. Hong, ``Cross-silo horizontal federated
  learning for flow-based time-related-features oriented traffic
  classification,'' in \emph{2020 21st Asia-Pacific Network Operations and
  Management Symposium (APNOMS)}, 2020, pp. 389--392.

\bibitem{emmert2020introductory}
F.~Emmert-Streib, Z.~Yang, H.~Feng, S.~Tripathi, and M.~Dehmer, ``An
  introductory review of deep learning for prediction models with big data,''
  \emph{Frontiers in Artificial Intelligence}, vol.~3, p.~4, February 2020.

\bibitem{pouyanfar2018survey}
S.~Pouyanfar, S.~Sadiq, Y.~Yan, H.~Tian, Y.~Tao, M.~P. Reyes, M.-L. Shyu, S.-C.
  Chen, and S.~S. Iyengar, ``A survey on deep learning: Algorithms, techniques,
  and applications,'' \emph{ACM Computing Surveys (CSUR)}, vol.~51, no.~5, pp.
  1--36, September 2018.

\bibitem{masdari2016overview}
M.~Masdari, S.~S. Nabavi, and V.~Ahmadi, ``An overview of virtual machine
  placement schemes in cloud computing,'' \emph{Journal of Network and Computer
  Applications}, vol.~66, pp. 106--127, May 2016.

\bibitem{thakur2015interoperability}
P.~Thakur and D.~K. Shrivastava, ``Interoperability issues and standard
  architecture for service delivery in federated cloud: a review,'' in
  \emph{International Conference on Computational Intelligence and
  Communication Networks}.\hskip 1em plus 0.5em minus 0.4em\relax IEEE, 2015,
  pp. 908--912.

\bibitem{atlam2018blockchain}
H.~F. Atlam, A.~Alenezi, M.~O. Alassafi, and G.~B. Wills, ``Blockchain with
  internet of things: Benefits, challenges, and future directions,''
  \emph{International Journal of Intelligent Systems \& Applications}, vol.~10,
  no.~6, 2018.

\bibitem{stoyanova2020survey}
M.~Stoyanova, Y.~Nikoloudakis, S.~Panagiotakis, E.~Pallis, and E.~K. Markakis,
  ``A survey on the internet of things (iot) forensics: challenges, approaches,
  and open issues,'' \emph{IEEE Communications Surveys \& Tutorials}, vol.~22,
  no.~2, pp. 1191--1221, Secondquarter 2020.

\end{thebibliography}

\begin{IEEEbiography}[{\includegraphics[width=1in,height=1.25in,clip,keepaspectratio]{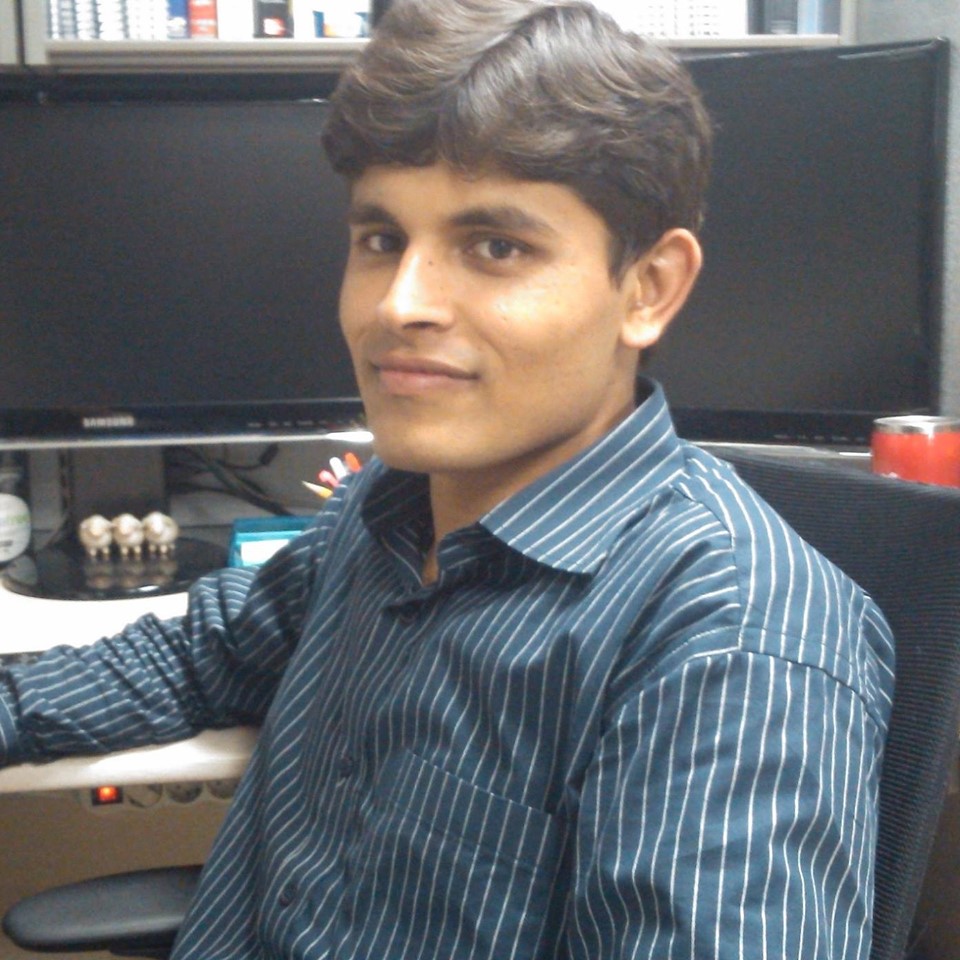}}]{Latif U. Khan} received his Ph.D. degree (Computer Engineering) and MS (Electrical Engineering) degree with distinction from Kyung Hee University (KHU), South Korea in 2021 and University of Engineering and Technology (UET), Peshawar, Pakistan, in 2017, respectively. He worked as a leading researcher in the intelligent Networking Laboratory under a project jointly funded by the prestigious Brain Korea 21st Century Plus and Ministry of Science and ICT, South Korea. Prior to joining the KHU, he has served as a faculty member and research associate in the UET, Peshawar, Pakistan. He has published his works in highly reputable conferences and journals. He is the author/co-author of two conference best paper awards. He is also author of two books, such as "Network Slicing for $5$G and Beyond Networks" and "Federated Learning for Wireless Networks". His research interests include analytical techniques of optimization and game theory to edge computing, end-to-end network slicing, and federated learning for wireless networks. 
\end{IEEEbiography}

\begin{IEEEbiography}[{\includegraphics[width=1in,height=1.25in,clip,keepaspectratio]{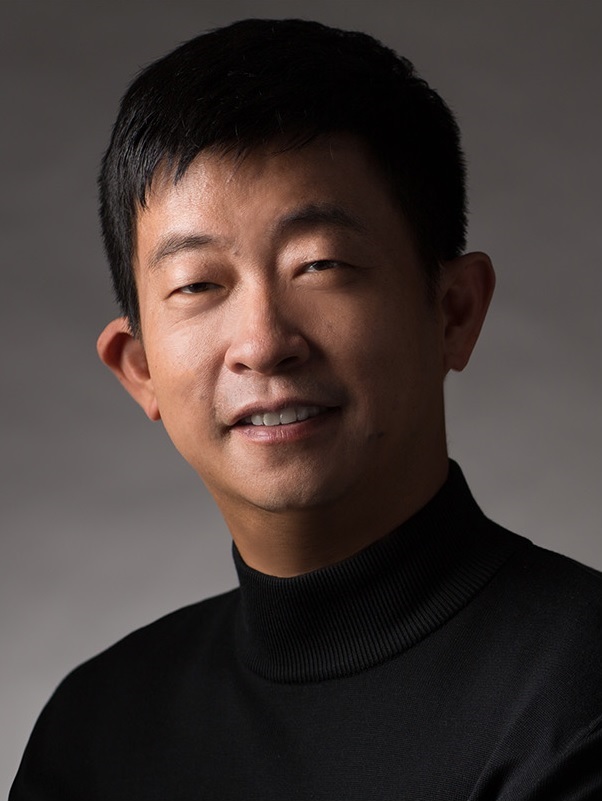}}]{Zhu Han}(S’01, M’04, SM’09, F’14) received the B.S. degree in electronic engineering from Tsinghua University, in 1997, and the M.S. and Ph.D. degrees in electrical and computer engineering from the University of Maryland, College Park, in 1999 and 2003, respectively. 

From 2000 to 2002, he was an R\&D Engineer of JDSU, Germantown, Maryland. From 2003 to 2006, he was a Research Associate at the University of Maryland. From 2006 to 2008, he was an assistant professor at Boise State University, Idaho. Currently, he is a John and Rebecca Moores Professor in the Electrical and Computer Engineering Department as well as in the Computer Science Department at the University of Houston, Texas. His research interests include wireless resource allocation and management, wireless communications and networking, game theory, big data analysis, security, and smart grid.  Dr. Han received an NSF Career Award in 2010, the Fred W. Ellersick Prize of the IEEE Communication Society in 2011, the EURASIP Best Paper Award for the Journal on Advances in Signal Processing in 2015, IEEE Leonard G. Abraham Prize in the field of Communications Systems (best paper award in IEEE JSAC) in 2016, and several best paper awards in IEEE conferences. Dr. Han was an IEEE Communications Society Distinguished Lecturer from 2015-2018, AAAS fellow since 2019 and ACM distinguished Member since 2019. Dr. Han is 1\% highly cited researcher since 2017 according to Web of Science. Dr. Han is also the winner of 2021 IEEE Kiyo Tomiyasu Award, for outstanding early to mid-career contributions to technologies holding the promise of innovative applications, with the following citation: ``for contributions to game theory and distributed management of autonomous communication networks."

\end{IEEEbiography}

\begin{IEEEbiography}[{\includegraphics[width=1in,height=1.25in,clip,keepaspectratio]{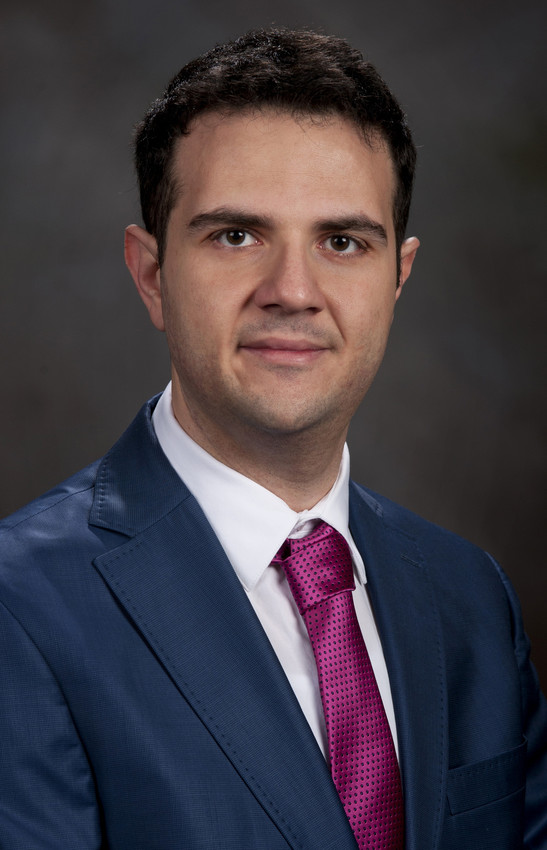}}]{Walid Saad } (S’07, M’10, SM’15, F’19) received the Ph.D. degree from the University of Oslo, Oslo, Norway, in 2010.,He is currently a Professor with the Department of Electrical and Computer Engineering, Virginia Tech, Blacksburg, VA, USA, where he leads the Network sciEnce, Wireless, and Security (NEWS) Laboratory. His research interests include wireless networks, machine learning, game theory, security, unmanned aerial vehicles (UAV), cyber-physical systems, and network science.,Dr. Saad was named the Stephen O. Lane Junior Faculty Fellow at Virginia Tech, from 2015 to 2017. In 2017, he was named as the College of Engineering Faculty Fellow. He was a recipient of the NSF CAREER Award in 2013, the Air Force Office of Scientific Research (AFOSR) Summer Faculty Fellowship in 2014, and the Young Investigator Award from the Office of Naval Research (ONR) in 2015. He was the author/coauthor of eight conference best paper awards at WiOpt in 2009, the International Conference on Internet Monitoring and Protection (ICIMP) in 2010, the IEEE Wireless Communications and Networking Conference (WCNC) in 2012, the IEEE International Symposium on Personal, Indoor and Mobile Radio Communications (PIMRC) in 2015, IEEE SmartGridComm in 2015, EuCNC in 2017, IEEE GLOBECOM in 2018, and the International Federation for Information Processing (IFIP) International Conference on New Technologies, Mobility and Security (NTMS) in 2019. He was also a recipient of the 2015 Fred W. Ellersick Prize from the IEEE Communications Society, the 2017 IEEE ComSoc Best Young Professional in Academia award, the 2018 IEEE ComSoc Radio Communications Committee Early Achievement Award, and the 2019 IEEE ComSoc Communication Theory Technical Committee. He received the Dean’s Award for Research Excellence from Virginia Tech in 2019. He also serves as an Editor for the IEEE Transactions on Wireless Communications, the IEEE Transactions on Mobile Computing, and the IEEE Transactions on Cognitive Communications and Networking. He is also an Editor-at-Large for the IEEE Transactions on Communications. He is also an IEEE Distinguished Lecturer. (Based on document published on 18 May 2020).
\end{IEEEbiography}

\begin{IEEEbiography}[{\includegraphics[width=1.5in,height=1.25in,clip,keepaspectratio]{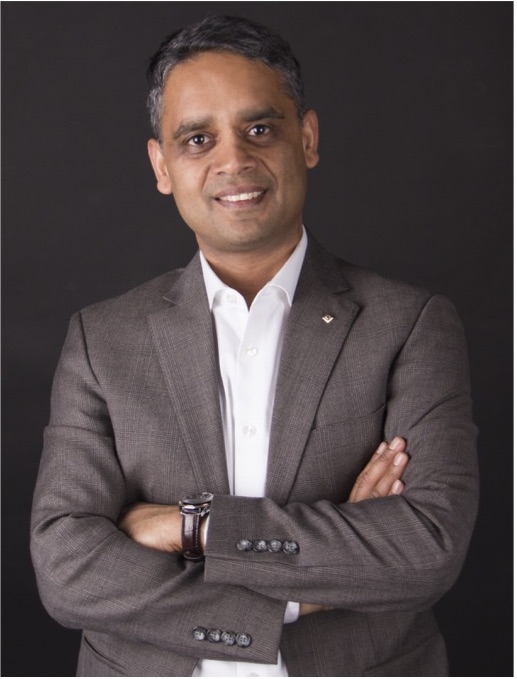}}]{Ekram Hossain}(F'15) is a Professor in the Department of Electrical and Computer Engineering at University of Manitoba, Canada. He is a Member (Class of 2016) of the College of the Royal Society of Canada, a Fellow of the Canadian Academy of Engineering, and a Fellow of the Engineering Institute of Canada.  He was elevated to an IEEE Fellow ``for contributions to spectrum management and resource allocation in cognitive and cellular radio networks''.  He was listed as a Clarivate Analytics Highly Cited Researcher in Computer Science in 2017, 2018, 2019, and 2020. Currently he serves as the Editor-in-Chief of IEEE Press. Previously he served as the Editor-in-Chief for the IEEE Communications Surveys and Tutorials (2012--2016).

\end{IEEEbiography}

\begin{IEEEbiography}[{\includegraphics[width=1.5in,height=1.25in,clip,keepaspectratio]{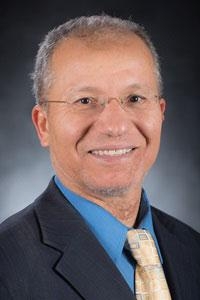}}]{Mohsen Guizani} (S’85, M'89, SM'99, F’09) received his B.S. (with distinction) and M.S. degrees in electrical engineering, and M.S. and Ph.D. degrees in computer engineering from Syracuse University, New York, in 1984, 1986, 1987, and 1990, respectively. He is currently a professor in the Computer Science and Engineering Department at Qatar University. Previously, he served in different academic and administrative positions at the University of Idaho, Western Michigan University, the University of West Florida, the University of Missouri-Kansas City, the University of Colorado-Boulder, and Syracuse University. His research interests include wireless communications and mobile computing, computer networks, mobile cloud computing, security, and smart grid. He is currently the Editor-in-Chief of IEEE Network, serves on the Editorial Boards of several international technical journals, and is the Founder and Editor-in-Chief of the Wireless Communications and Mobile Computing journal (Wiley). He is the author of nine books and more than 500 publications in refereed journals and conferences. He has guest edited a number of Special Issues in IEEE journals and magazines. He has also served as a TPC member, Chair, and General Chair of a number of international conferences. Throughout his career, he received three teaching awards and four research awards. He also received the 2017 IEEE Communications Society WTC Recognition Award as well as the 2018 AdHoc Technical Committee Recognition Award for his contribution to outstanding research in wireless communications and ad hoc sensor networks. He was the Chair of the IEEE Communications Society Wireless Technical Committee and the Chair of the TAOS Technical Committee. He served as a IEEE Computer Society Distinguished Speaker and is currently an IEEE ComSoc Distinguished Lecturer. He is a Senior Member of ACM.

\end{IEEEbiography}

\begin{IEEEbiography}[{\includegraphics[width=1in,height=1.25in,clip,keepaspectratio]{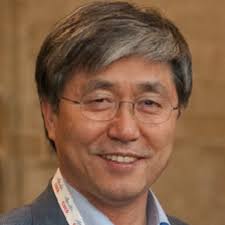}}]{Choong Seon Hong} (S’95-M’97-SM’11)  received the B.S. and M.S. degrees in electronic engineering from Kyung Hee University, Seoul, South Korea, in 1983 and 1985, respectively, and the Ph.D. degree from Keio University, Tokyo, Japan, in 1997. In 1988, he joined KT, Gyeonggi-do, South Korea, where he was involved in broadband networks as a member of the Technical Staff. Since 1993, he has been with Keio University. He was with the Telecommunications Network Laboratory, KT, as a Senior Member of Technical Staff and as the Director of the Networking Research Team until 1999. Since 1999, he has been a Professor with the Department of Computer Science and Engineering, Kyung Hee University. His research interests include future Internet, intelligent edge computing, network management, and network security. 
Dr. Hong is a member of the Association for Computing Machinery (ACM), the Institute of Electronics, Information and Communication Engineers (IEICE), the Information Processing Society of Japan (IPSJ), the Korean Institute of Information Scientists and Engineers (KIISE), the Korean Institute of Communications and Information Sciences (KICS), the Korean Information Processing Society (KIPS), and the Open Standards and ICT Association (OSIA). He has served as the General Chair, the TPC Chair/Member, or an Organizing Committee Member of international
conferences, such as the Network Operations and Management Symposium (NOMS), International Symposium on Integrated Network Management (IM), Asia-Pacific Network Operations and Management Symposium (APNOMS), End-to-End Monitoring Techniques and Services (E2EMON), IEEE Consumer Communications and Networking Conference (CCNC), Assurance in Distributed Systems and Networks (ADSN), International Conference on Parallel Processing (ICPP), Data Integration and Mining (DIM), World Conference on Information Security Applications (WISA), Broadband Convergence Network (BcN), Telecommunication Information Networking Architecture (TINA), International Symposium on Applications and the Internet (SAINT), and International Conference on Information Networking (ICOIN). He was an Associate Editor of the IEEE TRANSACTIONS ON NETWORK AND SERVICE MANAGEMENT and the IEEE JOURNAL OF COMMUNICATIONS AND NETWORKS. He currently serves as an Associate Editor for the International Journal of Network Management.
\end{IEEEbiography}

\end{document}